\documentclass[journal]{IEEEtran}
\usepackage{cite}
\usepackage{amsmath,amssymb,amsfonts}
\usepackage{algorithmic}
\usepackage{graphicx}
\usepackage{textcomp}
\usepackage{xcolor}
\usepackage{url}
\usepackage{multirow}
\usepackage{booktabs}
\usepackage{balance}
\usepackage{ltxtable}
\usepackage{lscape}
\usepackage{float}
\usepackage[linkcolor=black,anchorcolor=green,citecolor=blue]{hyperref}
\usepackage{color,colortbl}
\definecolor{grayone}{gray}{.8}
\usepackage[normalem]{ulem}
\def\BibTeX{{\rm B\kern-.05em{\sc i\kern-.025em b}\kern-.08em
    T\kern-.1667em\lower.7ex\hbox{E}\kern-.125emX}}
\usepackage{makecell}
\usepackage{threeparttable}
\makeatletter  
\newif\if@restonecol  
\renewcommand\footnoterule{%
	\kern-3\p@
	\hrule\@width\columnwidth
	\kern2.6\p@}

\newcommand{\sen}[1]{{\textcolor{black}{#1}}}

\newcommand{\revised}[1]{{\textcolor{black}{#1}}}

\newcommand{\updated}[1]{{\textcolor{black}{#1}}}

\newcommand{\tabincell}[2]{\begin{tabular}{@{}#1@{}}#2\end{tabular}}
\newcommand{\step}[1]{\textcircled{\small{#1}}}

\newcommand{\distance}{2pt}
\begin{document}
\title{A Performance-Sensitive Malware Detection System Using Deep Learning on Mobile Devices}

\author{Ruitao Feng,
Sen Chen\IEEEauthorrefmark{1},
Xiaofei Xie,
Guozhu Meng,
Shang-Wei Lin and
Yang Liu
\IEEEcompsocitemizethanks{
    \IEEEcompsocthanksitem \IEEEauthorrefmark{1}Sen Chen is the corresponding author. Email: chensen@ntu.edu.sg
	\IEEEcompsocthanksitem 
	Ruitao Feng, Xiaofei Xie, Shang-Wei Lin and Yang Liu are with the School of Computer Science and Engineering, Nanyang Technological University, Singapore. 
	Email: \{rtfeng, xfxie, shang-wei.lin, yangliu\}@ntu.edu.sg
	\IEEEcompsocthanksitem Sen Chen is with the College of Intelligence and Computing, Tianjin University, China and Nanyang Technological University, Singapore.
	\IEEEcompsocthanksitem Guozhu Meng is with the Institute of Information Engineering, Chinese Academy of Sciences, China.
	Email: mengguozhu@iie.ac.cn
    }
}

\markboth{Journal of IEEE Transactions on Information Forensics and Security,~Vol.~xx, No.~xx, xx~2020}%
{Feng \MakeLowercase{\textit{et al.}}: A Performance-Sensitive Malware Detection System Using Deep Learning on Mobile Devices}

\maketitle

\begin{abstract}
Currently, Android malware detection is mostly performed on server side against the increasing number of malware.
Powerful computing resource provides more exhaustive protection for app markets than maintaining detection by a single user.
However, apart from the applications (apps) provided by the official market (i.e., Google Play Store), apps from unofficial markets and third-party resources are always causing serious security threats to end-users.
Meanwhile, it is a time-consuming task if the app is downloaded first and then uploaded to the server side for detection, because the network transmission has a lot of overhead.
In addition, the uploading process also suffers from the security threats of attackers.
Consequently, a last line of defense on mobile devices is necessary and much-needed.

In this paper, we propose an effective Android malware detection system, MobiTive, leveraging customized deep neural networks to provide a real-time and responsive detection environment on mobile devices. 
MobiTive is a pre-installed solution rather than an app scanning and monitoring engine using after installation, which is more practical and secure.
Although a deep learning-based approach can be maintained on server side efficiently for malware detection, original deep learning models cannot be directly deployed and executed on mobile devices due to various performance limitations, such as computation power, memory size, and energy.
Therefore, we evaluate and investigate the following key points:
(1) the performance of different feature extraction methods based on source code or binary code;
(2) the performance of different feature type selections for deep learning on mobile devices;
(3) the detection accuracy of different deep neural networks on mobile devices; 
(4) the real-time detection performance and accuracy on different mobile devices; 
(5) the potential based on the evolution trend of mobile devices' specifications; and finally we further propose a practical solution (MobiTive) to detect Android malware on mobile devices.

\end{abstract}


\IEEEpeerreviewmaketitle

\section{Introduction}
\IEEEPARstart{W}{ith} the currently increasing number of Android devices and applications (apps), plenty of Android users are benefited from that.
\sen{The security and privacy concerns are also increasingly becoming the focus point to various mobile users and stakeholders. For example, more and more users store their personal data in mobile devices~\cite{chen2019ausera,chen2018mobile} through various popular apps such as shopping, banking, and social apps.
Consequently, since the last decade, attackers shift their attention to mobile apps. That makes Android malware undoubtedly become one of the most important security threats in this security field~\cite{zhou2012dissecting, tang2019large}.}

Therefore, how to detect Android malware becomes a severe problem.
End-users always expect a secure environment which is maintained by the app markets.
In other words, they consider their app sources are all trustable and secure enough.
It is not surprising that the demands of Android malware detection approaches have been proposed, such as signature-based approaches~\cite{zhou2012hey, zhou2013fast}, behavior-based approaches~\cite{yan2012droidscope, tam2015copperdroid}, data-flow analysis-based approaches~\cite{arzt2014flowdroid, li2015iccta}. We note that machine learning-based approach~\cite{arp2014drebin, yang2014droidminer,chen2016stormdroid,mariconti2016mamadroid,chen2016towards,fan2016poster,chen2018automated,yu2014towards} is one of the most promising techniques in detecting Android malware. 
With the available big data and hardware evolution over the past decade, deep learning has achieved tremendous success in many cutting-edge domains, including Android malware detection.
Actually, all of the above protecting solutions are mostly on server side for app markets.
However, when a new Android malware family is reported, not all the app markets are able to respond in a responsive time.
The current analysis workflow always follows analyzing malicious behaviors within apps, building the detection models with the generated features and then performing the detection on the entire apps.
Since the number of the real-world Android apps is extremely large, e.g., there are more than 3 million Android apps on Google Play Store, it is a time-consuming task to perform the complete detection with that large number of apps.
Moreover, the apps from unofficial markets and third-party resources like XDA~\cite{xda} are more vulnerable in the wild.
The security of these kinds of apps is indeed unpredictable and uncontrollable.

The traditional server-side based malware detection surely has unignorable drawbacks when detecting such apps, because
(1) it is a time-consuming task to upload the apps to server before the installation, especially for large apps;
(2) the uploading process via the Internet is not secure. 
For example, attackers may modify the malware during the uploading period such that an incorrect ``benign'' result is returned.
As a result, the users will install the malware.
Hence, a last line of defense on mobile devices is necessary and much-needed.
To address the severe problem, we intend to conduct Android malware detection on mobile devices instead of server side.

Actually, machine learning-based approaches have achieved better performance compared with other approaches in Android malware detection~\cite{arp2014drebin, chen2016stormdroid,yuan2016droiddetector,chen2018automated,wu2020xmal}.
In this paper, we intend to deploy the trained deep learning (DL) models from server-side to mobile devices.
While a computationally intensive deep learning software could be executed efficiently on server-side with the GPU support, such deep learning models usually cannot be directly deployed and executed on other platforms supported by small mobile devices due to various computation resource limitations such as the computation power, memory size, and energy.
In our previous work~\cite{feng2019mobidroid}, we leverage {TensorFlow Lite}~\cite{tensorflow-lite} to migrate the deep learning models. 
\revised{We proposed a convolutional neural network (CNN)-based Android malware detection system on mobile platform, which leveraged three kinds of features from decompiled Android apps according to the performance-based feature selection mechanism.
We have substantially extended our previous work from the following aspects:}

\begin{itemize}
    \item \sen{In the conference version~\cite{feng2019mobidroid}, we only focused on the performance of different feature types extracted from decompiled files such as smali files. To reach the best performance on mobile devices, we take the installation mechanism in the Android operation system into account. Specifically, we analyze and extract two types of features (i.e., manifest properties and API calls) 
    from {Dalvik binary files} directly instead of the decompiled files.}
    \item \sen{Meanwhile, to enrich the malicious behavior coverage of our selected features, we perform an empirical analysis to understand the existing malicious behaviors, most of which are collected from industrial malware analysis reports (e.g., Symantec Threats~\cite{symantec}). According to the understanding, we further update the feature inputs with the matching results between text-based behavior descriptions and code level features {(details on our website~\cite{mobitive}})}.
    \item \sen{To figure out the potential detection accuracy promotion of different deep neural networks, we not only apply our new extracted features with CNN models, but also present six more kinds of recurrent neural networks models (e.g., LSTM and GRU).
    Finally, we customized one RNN model to adopt the device-based detection scenario. \updated{Moreover, we further compare with four other existing Android malware detection approaches to demonstrate the effectiveness and efficiency of our approach.}}
    \item \updated{To investigate the effectiveness of our system on multi-class classification task, we demonstrate the result on classifying 701,300 Android malware into 21 families with our system.}
    \item \sen{To peek into the average usability and best practice for our new system, we evaluate our system on six real mobile devices from different manufacturers such as Google, Huawei, and Samsung, which released between 2015 and 2019. \updated{Meanwhile, we conduct a run-time performance evaluation with other device-end solution such as dynamic behavior analysis to demonstrate the effectiveness of our approach.} We also investigate the development trend of Android mobile phones to further understand the system usability.}
\end{itemize}

\sen{According to the evaluation metrics of accuracy and time cost from different features and neural networks, we propose an effective and efficient Android malware detection system on mobile devices, named MobiTive. MobiTive leverages
(1) a newly-proposed feature extraction method from binary code;
(2) a performance-based feature type selection mechanism;
(3) a novel feature updating method through malicious behavior mining and understanding;
(4) a customized deep neural network for classification.
So that, MobiTive can provide a real-time and fast responsive environment on mobile devices.} 

\sen{In our comprehensive experiments, (1) we first divide the feature preparation procedure into two steps, which are {raw data extraction} and {feature extraction}, and evaluate the performance (time cost) separately to decide the feature selection. 
(2) With the selected features, we then provide an accuracy comparison between different feature categories.
(3) The behavior-based feature updating method performs around {1\%$
\sim$5\%} accuracy increase.
(4) We provide a comprehensive comparison between seven different neural networks (e.g., CNN, LSTM, and GRU) to show the potential improvement of our customized DL models on network definition.
(5) \updated{We further evaluate the performance and accuracy of MobiTive on different real mobile devices by using our customized RNN model and compare with dynamic device-end solutions.}
(6) In the last part of our experiments, we perform an analysis of the performance trend on mobile devices from three different aspects and integrate the results to provide a strong evidence on the potential of MobiTive in practice.
Specifically, MobiTive achieves a relatively higher classification accuracy (i.e., 96.78\% accuracy) on real testing data in the wild and mobile devices with relatively lower overhead (i.e., less than 3 seconds on average for one app).} 

\revised{In summary, we make the following main contributions.}
\begin{itemize}
    \item \sen{We propose MobiTive, a device-end solution to protect mobile devices from malware threats in real-time efficiently by leveraging customized deep neural networks and binary features.
    \updated{This research work aims to detect malware directly on mobile devices as a pre-installed and run-time solution rather than detecting them on common servers or monitoring them after installation.}}
    \item \sen{We propose a new feature extraction method from binary code, as well as a feature updating method based on the understanding of malicious behaviors. \updated{Due to the high performance demand of mobile devices, we evaluate the different performance (time cost) and accuracy with various feature types and neural networks, and further provide a comparison against four existing Android malware detection approaches. Besides, we also investigate the accuracy on multi-class classification task.}}
    \item \sen{We evaluate and investigate the different performance on multiple devices from different manufacturers, and further provide insights of the current quality and potential for our approach according to the feature extraction and prediction time cost on six real mobile devices. \updated{Meanwhile, an additional comparison on run-time efficiency and discussion on effectiveness is provided to show the advantages against dynamic malware detection system based on behavior analysis.}}
\end{itemize}


\section{Preliminaries} \label{preliminaries}
\revised{In this section, we briefly introduce the structure of Android apps and Dalvik executable, the existing Android security mechanisms, and the migration/quantization procedure of trained DL models on PC/Server side.}

\subsection{Android Apps} \label{preliminaries:android_app}
To execute the code of Android apps, Android developers compile their source code and other components, like application structure files and other resources, etc., into an Android application package (APK). APK is a compressed application file for Android platform, which is used to deliver Android mobile applications. For each APK, it contains a manifest file (i.e., AndroidManifest.xml), Dex files, resources, assets, etc.

The manifest file contains the meta-data for Android apps, which defines the package name and application ID, app components like Intent filters, activities, and services, etc., permissions, device compatibility, like uses-feature and uses-sdk, etc.
Dex files as extension are Dalvik executable code, which can be executed on Dalvik virtual machine in Android OS and converted from Java bytecode via an alternative instruction set.
To make them more accessible, Dex files are often decompiled into {smali} files by reverse engineering, which contain the same meaning contents, but have a better syntax format before manual analysis.
However, the decompiling procedure will always cost considerable time.

\subsection{Dalvik Executable} \label{preliminaries:dalvik}
Dalvik executable file contains 21 kinds of contents, which can be mainly divided into metadata and program information, etc.
The metadata information of Dalvik executable file is provided by the header, checksum, signature, etc. After them, it follows with the size and offset values of program information, like class definition identifiers, method identifiers, type identifiers, string identifiers, etc. 
Besides the above program information, the map offset is also an important component. It provides concrete mappings between static information, like strings and method names, etc. With the given offsets, we can easily access the defined static program information without decompiling the binary executable into human readable format.
Other than the static information, we can also get the compiled code contents with the offset of each method.

\subsection{Security Mechanisms} \label{preliminaries:security}
The existing security mechanisms can be mainly divided into \updated{three categories, which are application market, Android OS platform, and device-end aspects in practice.}

From the aspect of Android market, the official market (i.e., Google Play Store) provides a security verification when the APK uploaded. For instance, Google provides protection backed by its machine learning algorithm. Some high-quality third-party markets also present security check for the uploaded applications. For example, ApkMirror~\cite{apkmirror} not only provides the signature verification, but also performs a protection service provided by GuardSquare. However, most of current security check service provided by third-party markets is very simple and limited. Some of them only contain a signature verification, which can be bypassed easily. Therefore, the users, who download applications from the third-party markets, have to install and use it at their own security risk.

On the mobile devices, there exists a lot of antivirus applications provided. \updated{The most famous security applications, like Avast and AVG, mainly provide their antivirus services by monitoring the privacy-sensitive components (e.g., run-time permission requests), and scanning the signature of suspicious apps with their local or on-cloud virus database.} 
Besides the protection from outside, Android OS also provides some strong built-in security mechanisms, like application sandbox, etc. Application sandbox mechanism provides an independent execution environment for every application. 
Hence, the attack from an application can only work on its own requested components. 
For instance, if Bluetooth permissions and actions liked activities are not required in the application, the attack can never access the functions provided by Bluetooth.

\begin{figure*}[t]
    \centering
    \includegraphics[width=0.75\textwidth]{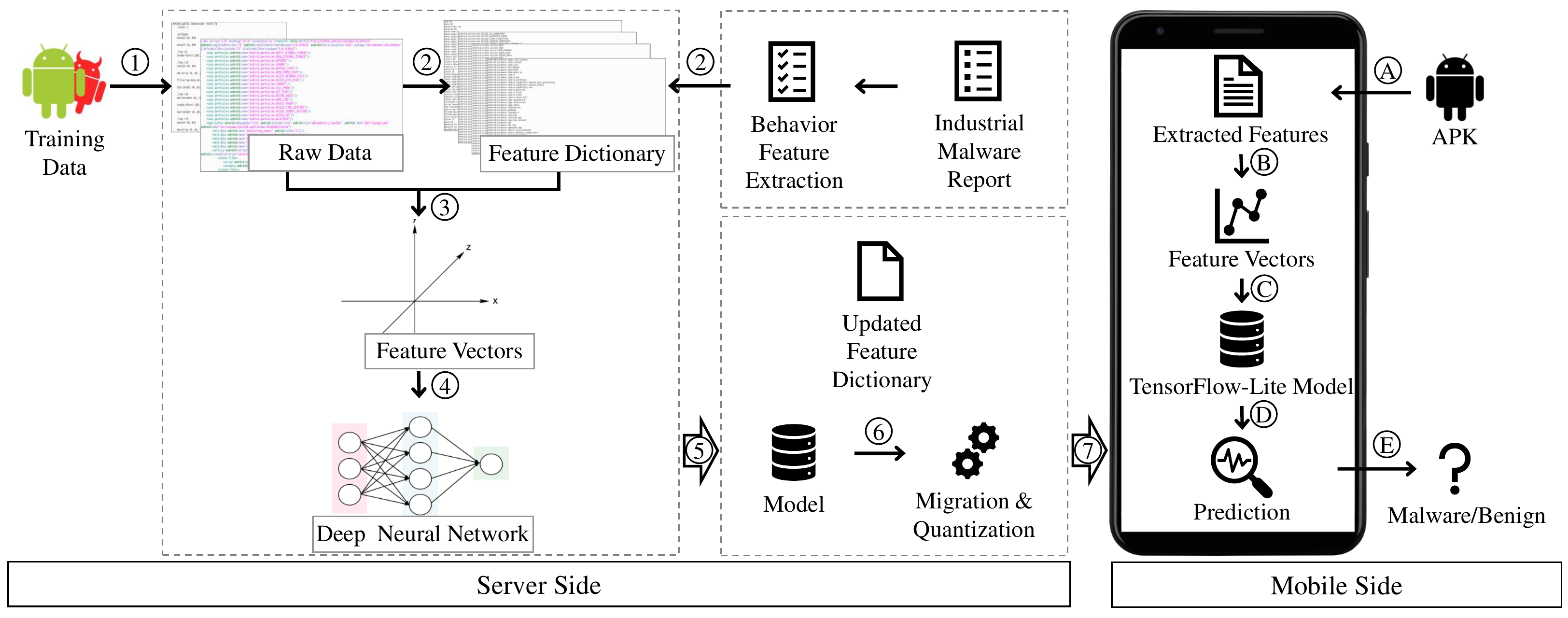}
    \caption{Overview of MobiTive}
    \label{fig:overview}
\end{figure*}

\subsection{Deep Learning Model Migration and Quantization} \label{preliminaries:migration_quantization}
After a DL model finishes the training process and is ready to deploy to a target device, it oftentime goes through either quantization, or platform migration, or both, before deployed to end-user applications.
This is because the training phase requires a vast amount of computation and energy resources. As the model size and the complexity of the tasks grow, more data are needed to train the network till reaching optimality, which could spend days, if not weeks, in training on high-performance GPU clusters. On the other hand, the deployment of the DNN models is usually faced with the resource-constrained environment with limited computation, power, etc.

Due to environment differences of a target platform~(e.g., mobile phones, {green energy embedded systems}) and training platform~(e.g., often equipped with GPUs), a DL model often goes through a customization phase to cater specific software and hardware constraints of a target platform. Quantization reduces the precision of a DL model so as to improve the computation efficiency, reduce memory consumption and storage size, which has become a common practice when migrating a large DL model trained on the cloud system to a mobile or IoT devices with low computation power.

Recently, the rapid development of system-on-chip (SoC) acceleration~(e.g., Qualcomm Snapdragon, Kirin 970, Samsung Exynos9) for AI applications provides the hardware support and foundation for universal deployment across platforms, especially on mobile device, edge computing device. Some lightweight solutions are proposed for mobile platforms such as {CoreML}~\cite{coreml}, {TensorFlow Lite}~\cite{tensorflow-lite}, {Caffe2} Mobile~\cite{caffe2-mobile} and PyTorch Android~\cite{pytorch-mobile}.
It proposes a chance to deploy the DL-based malware detection task on a mobile device directly.

\section{Approach} \label{approach}

\subsection{Overview of MobiTive} \label{approach:overview}
To achieve our target, we propose MobiTive, whose functionality could be divided into two main parts (i.e., parts of server side and mobile side), as shown in Fig.~\ref{fig:overview}. 
The first part of our system contains {feature preparation}, {DL model training}, {model migration and quantization}.
The second part is the deployment phase on mobile devices by using the migrated/quantized models.

\sen{In our previous work~\cite{feng2019mobidroid}, we involved multiple features (i.e., manifest properties, API calls, and opcode sequences) extracted from decompiled apps. In this paper, to improve the performance of MobiTive, we propose a new feature extraction method. Instead of decompiling APK into source code, like smali code, we extract and vectorize the manifest properties and API calls from binary code directly (step \step{1}).
We combine a performance-based feature selection mechanism and behavior-based feature updating method to generate the feature dictionary (step \step{2}).
With the customized deep neural networks and extracted feature vectors (steps \step{3} and \step{4}), the first part allows to provide a trained DL model and a feature dictionary for the second part (step \step{5}).
To make the model adaptive to mobile devices, we then migrate the pre-built DL model to a TensorFlow Lite model. 
Also, a quantization phase~\cite{quantization}, which is a general technique to reduce model size while also providing lower latency with little degradation in accuracy, is presented as a performance optimization for the mobile devices (step \step{6}).}

\sen{Fig.~\ref{fig:overview} shows that the second part loads the quantized DL model and feature dictionary into mobile devices.
After that, when an application is downloaded from market or third-party market, MobiTive can extract feature vectors from it and deliver the result to MobiTive (steps \step{A}$\rightarrow$\step{C}). 
After predicting with the loaded DL model, we obtain a certain level of confidence based on predictive output to know whether the downloaded Android app is a malware or not. (steps \step{D}$\rightarrow$\step{E}).} 

\subsection{Feature Preparation} \label{approach:feature_preparation}
To determine the features used in MobiTive, we perform a comparison of the extracting performance for most commonly-used features in previous malware detection approaches~\cite{arp2014drebin, narayanan2016adaptive, mclaughlin2017deep,chen2016stormdroid,chen2018automated}. 
Based on the performance-based feature selection method, manifest properties and API calls are selected in our device-end scenario (\emph{Feature Selection}).
Also, to update the feature dictionary and improve the representatives, we propose a behavior-based method based on industrial malware reports (\emph{Feature Updating}).
To get the features from APK, we unzip the package instead of decompiling it to reduce time cost.
Among the unzipped binary files, we can extract the features from raw data (\emph{Feature Extraction}).

\begin{table}\scriptsize
\caption{Selected Features}
\label{tab:selected_feature}
\centering
\begin{tabular}{ccc}
\hline
{\bf}    & {\bf \#API calls}  & {\bf \#Manifest properties}\\ \hline
\begin{tabular}[c]{@{}c@{}}Collected from samples or \\ by Android documentation\end{tabular}    & 2,989,011       & 613\\ \hline
Pruning by manual    & 1,509           & 613\\ \hline
\begin{tabular}[c]{@{}c@{}}Updated by\\ behavior-based analysis\end{tabular}
& 2,290           & 625\\ \hline
\end{tabular}
\begin{tablenotes}
\item Feature lists can be found on~\cite{mobitive}.
\end{tablenotes}
\end{table}

\subsubsection{Feature Selection} \label{approach:feature_preparation:selection}
\sen{Manifest properties such as used permissions, intents, and hardware features are widely-used features to detect Android malware~\cite{arp2014drebin, chen2016stormdroid,chen2018automated}. AndroidManifest.xml file can be easily decoded from APK file through existing tools, which benefits the feature extraction procedure.
It is belonging to a lightweight feature type, which would be adopted by the performance-sensitive system, like MobiTive.
In terms of the usefulness, API calls are more representative and important feature types because almost all malicious behaviors would be demonstrated by API callings.
Apart from the individual API call, API call sequence may contains more semantics, such as opcode code sequence.
However, the extraction procedure of these two feature types causes a lot of time due to analyzing source code or smali code. A novel feature extraction method for API calls is much-needed due to the energy limitation of mobile devices.
Besides the above two widely-used feature types, we also evaluate other potential structural features by their performance behaviors, such as inter-procedural control-flow graph (ICFG) and call graph (CG). ICFG not only provides the control-flow graph but also contains the inter-procedural between the components within apps. CG represents the calling relations between different methods.
By evaluating the performance (time cost) of all these potential feature types based on the two different extraction steps, which are raw data and feature vector extraction (steps \step{1} and \step{3}), we select unzipping APK to extract raw data first and select API calls and manifest properties as our feature types due to the better extraction performance compared with others (details in \S\ref{experiment:feature_selection}).}

\sen{To get the feature vectors (step \step{3}), we build a feature dictionary (step \step{2}) according to the two types of features selected by performance comparison.
Specifically, we build the manifest property dictionary by following the official Android documentation.
As shown in Table~\ref{tab:selected_feature}, the manifest properties contain 613 features in total, including 324 used permissions, 213 intents, and 76 hardware features.
In terms of the API call dictionary, we conduct a data-driven analysis to determine the feature lists.
Specifically, by parsing the API calls from more than 60k real-world Android apps collected from Google Play Store and malware, we collect 2,989,011 unique API calls in total.
We summarize three rules to reduce the size of API calls through manual analysis. Firstly, we remove the obfuscated API calls. Secondly, we delete the API calls that are not related to potential malicious behaviors, such as View loading API. Last, we remove the third-party API calls, because these API calls exist  and customize in an app, may rarely appear in other apps.
As shown in Table~\ref{tab:selected_feature}, after pruning, the number of selected API calls is only 1,509.
The details of feature lists can be found on our website~\cite{mobitive}. We build a feature dictionary based on the 1,509 API calls and 613 manifest properties for matching on the features of permission, intent, hardware, and API calls (step \step{2}).}

\subsubsection{Feature Updating} \label{approach:feature_preparation:updating}
\sen{The quality of machine learning-based detection approaches highly depends on the selected features, which means that a more comprehensive feature coverage of malicious behaviors makes more benefits MobiTive.
To enrich the feature coverage of malicious behaviors, we collect hundreds of industrial malware reports from Symantec Threats~\cite{symantec}.
With the collected text-based reports, we perform a manual analysis and summarize 23 kinds of basic potential malicious behaviors as a supplement for the selected features. Note that, the malware reports detail the malicious behaviors and the core code level features, including both API calls and manifest properties.
Also, a behavior-based feature understanding and verification by three co-authors are performed to ensure the manual results.
As a result, except the features in the original feature dictionary, there are 46 new API calls and 12 new manifest properties in total, which are updated for a new feature dictionary.
We also extend the new API calls with their package name. 
For example, if a new API call has package name as ``android/net/Uri'', we extract all the API calls under this package.
As shown in Table~\ref{tab:selected_feature}, there are 781 API calls extended according to the 46 new API calls.
Finally, we supplement our feature dictionary and update to 2,290 API calls and 625 manifest properties. The new feature dictionary is used to get the feature vector of each app for model training.}

\subsubsection{Feature Vector Extraction} \label{approach:feature_preparation:extraction}
As we mentioned in feature selection, the traditional feature vector extraction methods cause a lot of time due to the cost of decompiling and extracting from source code such as Java code and smali code. To improve the extraction performance, we propose a novel feature vector extraction method from binary code instead of source code.
Specifically, by analyzing the inner architecture of Dalvik binary file (classes.dex), we find there exists an API table, which is used to match the executable symbols and API strings.
We extract API calls by parsing the API table in classes.dex file based on the address and offset defined in the metadata.
Meanwhile, to get access to the information in binary format AndroidManifest.xml, we firstly generate a standard output with a XML decoder, Axmldec~\cite{axmldec}. By analyzing the decoded manifest file, the manifest properties can be extracted.

\subsection{DL Model Construction} \label{approach:dl_model}
\subsubsection{DL Model Training} \label{approach:dl_model:training}

\begin{table}\scriptsize
\caption{Deep Neural Network Architecture: GRU and LSTM}
\label{tab:RNN}
\centering
\sen{\begin{tabular}{c|c|c}
\hline
\multicolumn{3}{c}{{Input}}\\ \hline
\multicolumn{1}{c|}{\multirow{2}{*}{{Reshape}}} & \multicolumn{1}{c|}{input} & (None, 2915, ) \\ \cline{2-3} 
\multicolumn{1}{c|}{} & \multicolumn{1}{c|}{output} & (None, 1, 2915) \\ \hline
\multicolumn{1}{c|}{\multirow{2}{*}{{GRU/LSTM}}} & \multicolumn{1}{c|}{input} & (None, 1, 2915) \\ \cline{2-3} 
\multicolumn{1}{c|}{} & \multicolumn{1}{c|}{output} & (None, 128) \\ \hline
\multicolumn{3}{c}{{Dropout}} \\ \hline
\multicolumn{3}{c}{{Softmax Classification}} \\ \hline
\end{tabular}}
\end{table}

\updated{To discover the potential accuracy improvement and usability for different deep neural networks, we present seven widely-used networks to train the classifier, with the input feature vectors generated by step~\step{3}.}
As shown in Table~\ref{tab:RNN},~\ref{tab:Stacked_RNN} and ~\ref{tab:Bidirect_RNN}, we customize the RNN models to adopt the device-end scenario and improve performance.
For simple RNNs in Table~\ref{tab:RNN}, the first computational layer is a LSTM/GRU layer with 128 neural units. After the computation, the dimension of input tensor will reduce to 128 from (1, 2,915). Then, there will be a dropout layer, the dropout rate is 0.5.
At last, the result is passed to a softmax classifier function to get the final training result.
For stacked RNNs in Table~\ref{tab:Stacked_RNN}, there will two stacked LSTM/GRU with dropout layers instead.
For bidirectional RNNs in Table~\ref{tab:Bidirect_RNN}, we apply a bidirectional LSTM/GRU layer instead of the original LSTM/GRU layer.

Moreover, we build the convolutional neural network (CNN) with reference to the conference version~\cite{feng2019mobidroid}.
As shown in Table~\ref{tab:CNN}, the first layer of the CNN model is Zero Padding Layer.
With input feature vectors, we need to fit it to the training part. 
Hence, we add two nonsense dimensions to the end of input since the kernel size of our convolutional layer is $3\time3$.
Then, the resulting vector is reshaped to a matrix, whose horizontal dimension is 3, and send to the next layer.
The second layer is the convolution layer with a $3\time3$ kernel, which receives the embedded matrix as its input and applies the convolution filter to produce activation maps for each batch.
Before delivering the batches to the hidden layer, a global max pooling is used after activation to reduce the dimensions.
Finally, the vector is passed to a hidden full layer, which is a multi-layer perception, for classification. 
To detect the relation between the result vector, we construct two sub-layers in the hidden layer, each of them contains a Rectified Linear Unit activation function.
At last, the result from the hidden layer is passed to a softmax classifier function to get the final training result.

\begin{table}\scriptsize
\caption{Deep Neural Network Architecture: Stacked GRU and LSTM}
\label{tab:Stacked_RNN}
\centering
\sen{\begin{tabular}{c|c|c}
\hline
\multicolumn{3}{c}{{Input}}\\ \hline
\multicolumn{1}{c|}{\multirow{2}{*}{{Reshape}}} & \multicolumn{1}{c|}{input} & (None, 2915, ) \\ \cline{2-3} 
\multicolumn{1}{c|}{} & \multicolumn{1}{c|}{output} & (None, 1, 2915) \\ \hline
\multicolumn{1}{c|}{\multirow{2}{*}{{GRU/LSTM}}} & \multicolumn{1}{c|}{input} & (None, 1, 2915) \\ \cline{2-3} 
\multicolumn{1}{c|}{} & \multicolumn{1}{c|}{output} & (None, 1, 128) \\ \hline
\multicolumn{3}{c}{{Dropout}} \\ \hline
\multicolumn{1}{c|}{\multirow{2}{*}{{GRU/LSTM}}} & \multicolumn{1}{c|}{input} & (None, 1, 128) \\ \cline{2-3} 
\multicolumn{1}{c|}{} & \multicolumn{1}{c|}{output} & (None, 128) \\ \hline
\multicolumn{3}{c}{{Dropout}} \\ \hline
\multicolumn{3}{c}{{Softmax Classification}} \\ \hline
\end{tabular}}
\end{table}

\begin{table}\scriptsize
\caption{Deep Neural Network Architecture: Bidirectional GRU and LSTM}
\label{tab:Bidirect_RNN}
\centering
\sen{\begin{tabular}{c|c|c}
\hline
\multicolumn{3}{c}{{Input}}\\ \hline
\multicolumn{1}{c|}{\multirow{2}{*}{{Reshape}}} & \multicolumn{1}{c|}{input} & (None, 2915, ) \\ \cline{2-3} 
\multicolumn{1}{c|}{} & \multicolumn{1}{c|}{output} & (None, 1, 2915) \\ \hline
\multicolumn{1}{c|}{\multirow{2}{*}{{Bidirectional (GRU/LSTM)}}} & \multicolumn{1}{c|}{input} & (None, 1, 2915) \\ \cline{2-3} 
\multicolumn{1}{c|}{} & \multicolumn{1}{c|}{output} & (None, 256) \\ \hline
\multicolumn{3}{c}{{Dropout}} \\ \hline
\multicolumn{3}{c}{{Softmax Classification}} \\ \hline
\end{tabular}}
\end{table}

\begin{table}[t]\scriptsize
\caption{Deep Neural Network Architecture: CNN}
\label{tab:CNN}
\centering
\begin{tabular}{c|c|c}
\hline
\multicolumn{3}{c}{{Input}}\\ \hline
\multicolumn{1}{c|}{\multirow{2}{*}{{Reshape}}} & \multicolumn{1}{c|}{input} & (None, 2915, ) \\ \cline{2-3} 
\multicolumn{1}{c|}{} & \multicolumn{1}{c|}{output} & (None, 1, 2915) \\ \hline
\multicolumn{1}{c|}{\multirow{2}{*}{{Zero Padding Layer}}} & \multicolumn{1}{c|}{input} & (None, 1, 2915) \\ \cline{2-3} 
\multicolumn{1}{c|}{} & \multicolumn{1}{c|}{output} & (None, 1, 2916) \\ \hline
\multicolumn{1}{c|}{\multirow{2}{*}{{Reshape}}} & \multicolumn{1}{c|}{input} & (None, 1, 2916) \\ \cline{2-3} 
\multicolumn{1}{c|}{} & \multicolumn{1}{c|}{output} & (None, 3, 708) \\ \hline
\multicolumn{1}{c|}{\multirow{2}{*}{{Convolutional Layer}}} & \multicolumn{1}{c|}{input} & (None, 3, 708) \\ \cline{2-3} 
\multicolumn{1}{c|}{} & \multicolumn{1}{c|}{output} & (None, 64, 706) \\ \hline
\multicolumn{3}{c}{{Relu}} \\ \hline
\multicolumn{1}{c|}{\multirow{2}{*}{{Global Max Pooling}}} & \multicolumn{1}{c|}{input} & (None, 64, 706) \\ \cline{2-3} 
\multicolumn{1}{c|}{} & \multicolumn{1}{c|}{output} & (None, 64) \\ \hline
\multicolumn{1}{c|}{\multirow{2}{*}{{Linear Dense Layer}}} & \multicolumn{1}{c|}{input} & (None, 64) \\ \cline{2-3} 
\multicolumn{1}{c|}{} & \multicolumn{1}{c|}{output} & (None, 16) \\ \hline
\multicolumn{3}{c}{{Relu}} \\ \hline
\multicolumn{1}{c|}{\multirow{2}{*}{{Linear Dense Layer}}} & \multicolumn{1}{c|}{input} & (None, 16) \\ \cline{2-3} 
\multicolumn{1}{c|}{} & \multicolumn{1}{c|}{output} & (None, 2) \\ \hline
\multicolumn{3}{c}{{Softmax Classification}} \\ \hline
\end{tabular}
\end{table}

\subsubsection{DL Model Migration and Quantization} \label{approach:dl_model:migration_quantization}
\sen{To deploy our pre-trained DL model on mobile devices, we convert and migrate the pre-trained model to a TensorFlow Lite model, which is supported by Android operating system (step \step{6}).
Specifically, we migrate the TensorFlow model to a mobile readable TensorFlow Lite model with a TensorFlow Lite converter~\cite{tensorflow-lite}.
Apart from the model migration, we also quantize our pre-trained model to improve the performance on mobile devices, which does not affect the accuracy of detection much.
In the experiments, we measure the performance of accuracy and time cost affected by the model migration and quantization (details in \S \ref{experiment:mobile:acc_time_mobile}).}

\updated{\subsection{Real-Time Detection System}} \label{approach:mobitive}
{Before conducting a real-time detection, the quantized TensorFlow Lite model and feature dictionary should be deployed to the detection system in advance (step \step{7}). 
There are three main steps before completing the prediction.
The first step of MobiTive is feature preparation. 
When an APK file is received in step \step{A}, MobiTive first unzips it into original assembly files such as AndroidManifest.xml, classes.dex, and other resources.
Features of API calls and manifest properties will be extracted accordingly.
We implement an API parser to extract the API calls from classes.dex directly based on the understanding of Dalvik binary code. 
Since the raw binary AndroidManifest.xml cannot be analyzed directly, we use a third-party decoder library, AXML~\cite{AXML}, to get the decoded manifest file. By analyzing the decoded manifest file, the three kinds of manifest properties will be extracted from the XML tag.
Hence, we can get both the manifest property vector and API call vector in step \step{B}.
All the two types of features are transformed into a vector, we connect them together as the input of TensorFlow Lite model (step \step{C}).
With the quantized model, MobiTive can perform the prediction in step \step{D} and show the final prediction result as a feedback in step \step{E}.
\updated{With the prediction result, the system can raise a warning to help users blocking the installation of the detected malware and further save its information (e.g., name, version, checksum) in local database. Also, besides the actions on local devices, reporting the malicious applications' information to the corresponding market and synchronizing the malware information to the updating server can be another two options.}}

To deploy an update for the MobiTive in practice, the service provider firstly need to collect the new detected malware and update the training dataset. After updating, it will be able to obtain a new pre-trained model on server. Then, the new model can be packed as a system patch and deployed to devices within an update directly. As a result, the updated system surely will improve the effectiveness and robustness of the protection on device based on the new delivered model.

\noindent{\bf More implementation details.}
\updated{The AXML version used in MobiTive is v1.0.1.
The API parser used in MobiTive on Android devices is implemented based on the Dex2jar~\cite{dex2jar} (2.1-nightly-28). Unlike the original Dex2jar project, we do not decompile the Dalvik executable files (i.e., .dex files) back into .smali files or .class files. Instead, we only involve the binary formatting functions in Dex2jar and collect the API calls from the decoded API table. The API parser is served as an external lib file in the MobiTive.
Technically, the classification functionality of MobiTive on Android devices is consist of 3 main parts. Different from the well established high level API provided in Keras (2.2.4), the basic data structure used in the computation with TensorFlow Lite (0.0.0-nightly) on Android devices is bytebuffer. Thus, firstly, there will be a step to convert the input vector and model into bytebuffer format. Secondly, by loading the model into a TensorFlow Lite interpreter, we can feed the input bytebuffer into the interpreter and get the result matrix. At last, by using an argmax function on the result matrix, the final prediction result can be obtained.}

\section{Experiments} \label{experiment}
\updated{In this section, our experiments are technically organized into three subsections based on the model deployment environments (i.e., PC/server and mobile).
First, the goals of our experiments on PC/server are to investigate: 
(1) the performance of extraction time of different raw data (techniques) and feature types;
(2) the effectiveness of behavior-based feature updating method;
(3) the detection accuracy of different deep neural networks;
(4) the comparison with other existing learning-based Android malware detection solutions;
(5) the accuracy of multi-class classification on malware families.}

\updated{Second, based on the observed findings and obtained results, we further evaluate:
(1) the performance of feature preparation on six different real devices with six different app sizes from 5MB to 50MB;
(2) the efficiency of detecting with different RNN models on real devices;
(3) the usability (i.e., performance and accuracy) of MobiTive on six different real mobile devices;
(4) the efficiency of MobiTive by comparing to dynamic behavior-based run-time detection systems.}

\updated{In the end, we conduct a study on the hardware performance trend of Android mobile devices to provide insights into the future usability of MobiTive.}

\subsection{Experiment Environment}\label{experiment:dataset_environment:environment}
\sen{The experiments on server side are run on a Ubuntu 16.04 server with two Intel Xeon E5-2699 V3 CPUs, 192GB RAM, and NVIDIA GeForce 2080Ti GPU.
{To evaluate our approach, we select 6 different Android mobile devices to evaluate the performance and accuracy of our approach on real mobile devices. Among them, there are four common specification devices (Nexus 6P, Huawei Mate 10, HTC U11, and LG G6), a flagship device (Huawei P30), and a low-profile device (Samsung Galaxy J7 Pro) (detailed specifications provided on our website~\cite{mobitive}).}
The implementation language of our system on server is Python 3.
{To get access to the raw data and features, we use seven different kinds of existing tools, which are axmldec~\cite{axmldec}, AXML~\cite{AXML}, ApkTool~\cite{apktool}, AndroGuard~\cite{androguard}, Dex2jar~\cite{dex2jar}, Soot~\cite{soot}, and FlowDroid~\cite{arzt2014flowdroid}.}
{axmldec} is a C++ project which can be used to decode binary manifest file into readable XML format file. {AXML} is a library designed to parse binary Android XML files. 
It is written in Java and can be used in an Android app as an external library. 
{ApkTool} is a tool for reverse engineering, which can decompile the apk file and generate the resources, which contains manifest, {smali} files, and etc.
{AndroGuard} is a Python tool, which cannot only decode the resources but also disassemble bytecode to Java code. 
Also, with the help of {AndroGuard}, we can easily generate the call graphs (CG) and data-flow graph for an Android app. 
{Dex2jar} is a project which contains tools to work with Android .dex and Java .class files. 
{Soot} is a Java optimization framework, which can be used to extract the call graph (CG).
{FlowDroid} is a static taint analysis tool for Android apps. It is applied to generate inter-procedural control-flow graph (ICFG).
Apart from the above existing tools for feature extraction, 
{{JitPack} is a novel package repository for JVM and Android projects, which can build the project to a ready-to-use artifacts (i.e., jar and aar).}
The deep neural networks and training projects are implemented with Keras~\cite{keras}, Numpy, Scikit-learn~\cite{scikit}, and TensorFlow libraries~\cite{tensorflow}.} 

\subsection{\updated{Effectiveness of Feature Extraction, Feature Updating, Feature Category Selection, and Neural Network Selection}} \label{experiment:feature_selection}
\subsubsection{Performance evaluation of feature extraction} \label{experiment:feature_selection:raw_data_feature_extraction}
\updated{In this experiment, we split the feature extraction time into two parts respectively (i.e., APK$\rightarrow$raw data$\rightarrow$feature) along with the technical procedures in feature preparation phase to show the performance advantages of our selected features.}

\paragraph{Dataset} \label{experiment:feature_selection:raw_data_feature_extraction:dataset}
\updated{To mitigate the uncertain influence from apps' size on the time cost of feature extraction, we randomly collect 60 Android apps among 6 different sizes (i.e., 5MB, 10MB, 20MB, 30MB, 40MB, and 50MB) to provide a clear performance comparison between different extraction methods such as feature extraction from source code and binary code.}

\paragraph{Setup} \label{experiment:feature_selection:raw_data_feature_extraction:setup}
\updated{In this experiment, we first evaluate the extraction time (APK$\rightarrow$raw data) of 3 different raw data types, which are widely-used in the existing static analysis based malware detection work (i.e., ICFG extracted by FlowDroid, CG extracted by Soot and AndroGuard, decompiled files obtained by ApkTool), together with our selected extracting method (i.e., binary code obtained by unzipping).}

\updated{Secondly, apart from the above raw data extracting methods, we further evaluate the extracting performance (raw data$\rightarrow$feature) of 3 different feature types (i.e., manifest properties, API calls, and opcode sequence~\cite{feng2020seq}) that generated from two kinds of raw data types (i.e., decompiling and unzipping).
We do not further evaluate the graph-related features due to the large time cost of raw data extracting.}

\begin{figure}
    \includegraphics[width=0.45\textwidth]{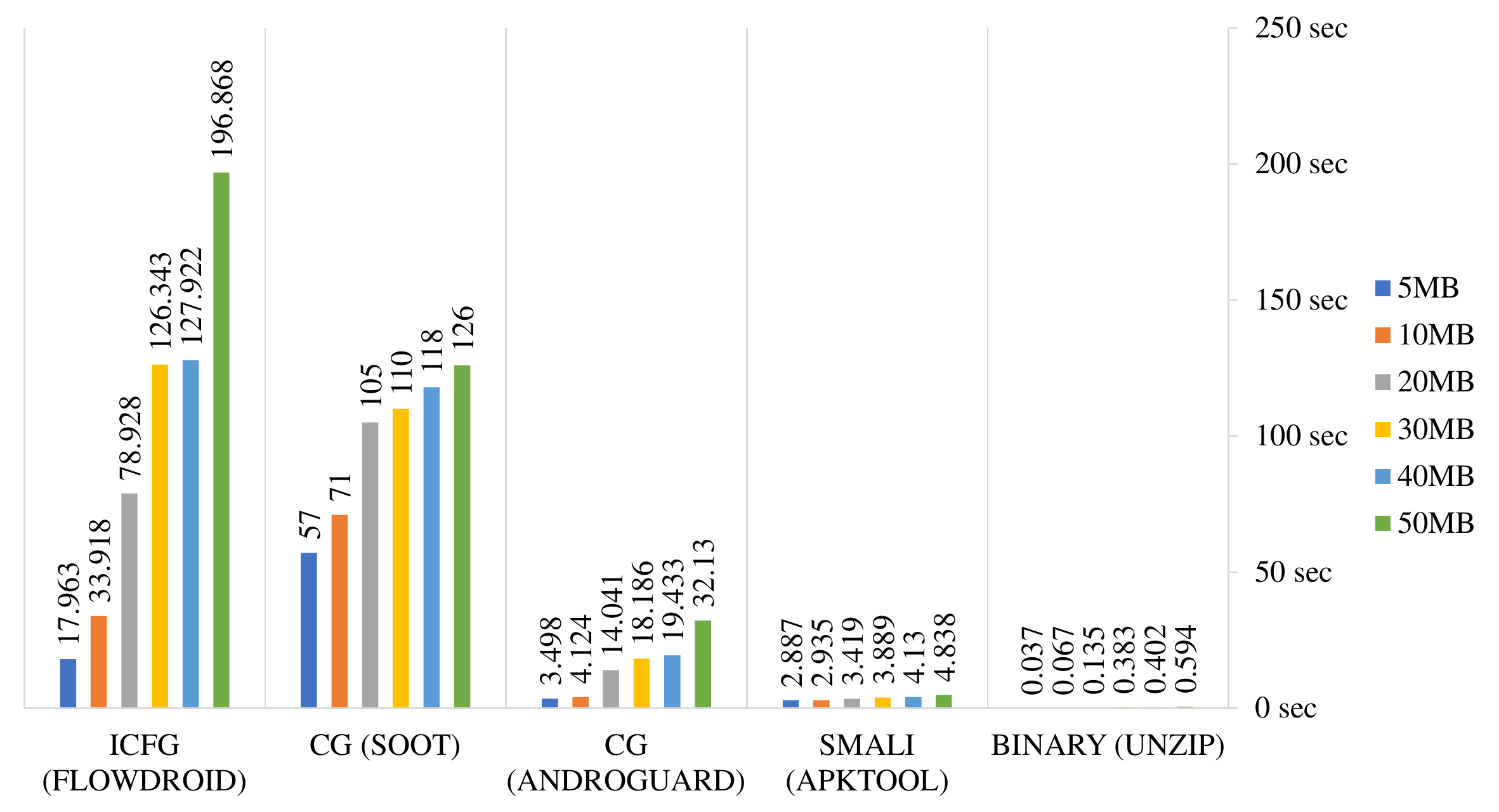}
    \centering
    \caption{\sen{Extraction time of different raw data types}}
    \label{fig:time_raw_data}
\end{figure}

\updated{For the decompiled manifest and smali files, we use a XML tag parser to extract manifest properties from manifest file decompiled by ApkTool. To extract API calls, we obtain the result by matching the API call dictionary and smali files directly. 
We extract the opcode sequences for each smali file by matching it to the opcode list~\cite{opcode}. 
For the unzipped binary manifest and Dalvik binary files, we evaluate the extraction time of 2 different feature types (i.e., manifest properties and API calls). 
To extract manifest properties from the binary manifest file, we apply Axmldec to extract manifest properties.
We extract API calls by loading the API table directly with the offset and size defined in the metadata of the Dalvik binary file.}

\paragraph{Results} \label{experiment:feature_selection:raw_data_feature_extraction:results}
\updated{We demonstrate the results from the 2 aspects (APK$\rightarrow$raw data and raw data$\rightarrow$feature) as below.}

\noindent{(1) \textit{Raw data extraction.}}
{Fig.~\ref{fig:time_raw_data} shows the extraction time of ICFG and CG is too large to be accepted performance-sensitive approach like MobiTive.
Specifically, extracting ICG via FlowDroid takes 196.868 seconds on 50MB apps and even 17.963 seconds on 5MB apps on average.
Generating CG with Soot takes 126 seconds on 50MB apps and 57 seconds on 5MB apps, and it costs 32.13 seconds and 3.498 seconds accordingly when prepared with AndroGuard. AndroGuard achieves a better performance than Soot on CG extraction.
As for MobiTive, the detection should be performed in a responsive period comparing to the app installing time, users cannot buy it if the reacting time takes too long.
The extraction time costs of decompiling, which applied in our previous work~\cite{feng2019mobidroid}, is acceptable but also limited, by comparing with the app installing time on average.
Considering the time cost of extracting raw data by unzipping, 5MB apps take only 0.037 seconds and 50MB apps take 0.594 seconds, which reaches a much better performance than the processing time of decompiling.
Therefore, we decide to use unzipping as our raw data extraction method.}

\begin{figure}
    \includegraphics[width=0.45\textwidth]{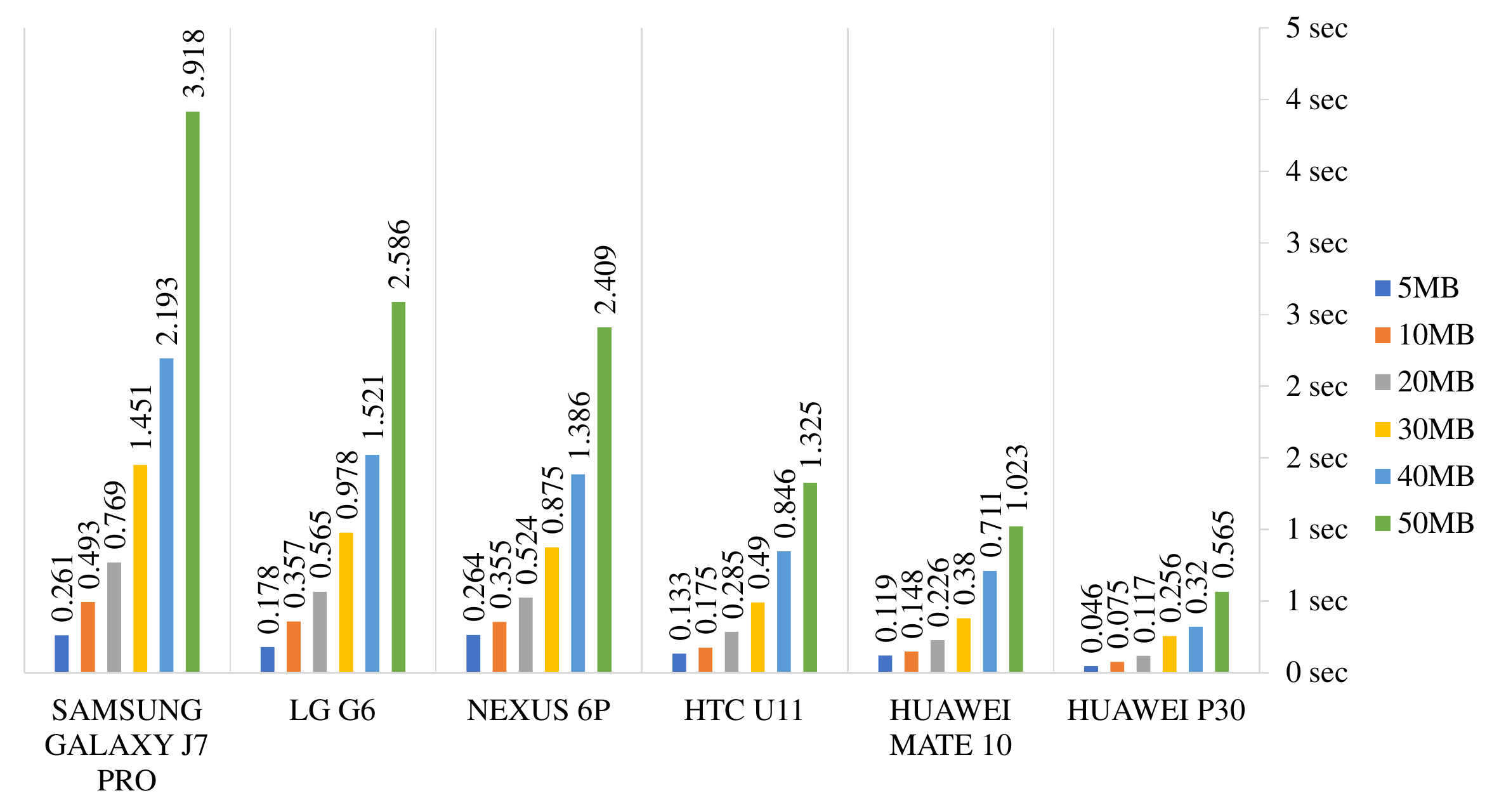}
    \centering
    \caption{\sen{Extraction time of different feature types}}
    \label{fig:time_feature_type}
\end{figure}

\noindent{(2) \textit{Feature extraction.}}
{Fig.~\ref{fig:time_feature_type} shows that the time consumption of features extracted from decompiled files is much longer than the same features generated directly from unzipped binary files.
Specifically, in terms of the time cost of API calls, extracting them from 5MB apps only takes only 0.042 seconds on average. However, if we extract the API calls from decompiled smali files, it takes 2.923 seconds.
For the 50MB apps, it will cost 0.601 and 5.002 seconds for extracting the API calls.
Considering the extraction time of manifest properties, 5MB and 50MB apps will take 2.89 and 4.841 seconds, when we extract the manifest properties from the decoded manifest file by ApkTool.
When we extract them from the unzipped binary manifest file, the time is reduced to 0.041 and 0.599.
Apart from manifest properties and API calls, we find that the extraction time of opcode sequence is much larger than the other two feature types. For 50MB apps, it will take over 6 seconds on average.
Therefore, to improve the performance of MobiTive in feature extraction, we decide to use the two feature types with shorter extraction time as our model inputs.}

\subsubsection{Accuracy evaluation of behavior-based feature updating method}
\label{experiment:feature_selection:updating}
\updated{In this experiment, we evaluate the effectiveness of the behavior-based feature updating method presented in \S\ref{approach:feature_preparation:updating} by comparing the results between the features used in our previous work~\cite{feng2019mobidroid} (MobiDroid) and MobiTive.}

\paragraph{Dataset} \label{experiment:feature_selection:updating:dataset}
{As shown in Table~\ref{tab:dataset}, we collect more than 70k Android apps in total as our evaluation subject. Specifically, these apps consist of 29,010 Android malware, and others are benign apps crawled from Google Play Store. However, these might be malware on the official market.
To filter the potential malware as far as possible, we upload them to VirusTotal~\cite{virustotal}, which is an online antivirus service with over 60 security scanners, to make a verification.
The 29,010 malicious samples contain 5,560 apps that downloaded from Drebin~\cite{arp2014drebin}, 1,260 apps validated in Genome project~\cite{zhou2012dissecting}, 20,000 crawled from VirusShare, and the remaining are used in KuafuDet~\cite{chen2018automated}, including 360 from Contagio Mobile Website~\cite{contagio} and 1,830 from Pwnzen Infotech Inc.~\cite{pwnzen}.
In summary, we collect a large-scale dataset of benign and malicious samples for the following experiments.
Since our dataset comes from multiple sources, there have a lot of duplicated samples. 
Therefore, we perform a hash check for eliminating redundant apps among malicious and benign apps. 
During the data prepossessing, which has raw data decompiling and feature vector generation steps, we receive some failed cases due to the capabilities of API parser.} 
\updated{The rest of the failures are just caused by the broken APK packages, we also remove them directly. As a result, we choose 18,000 benign and malicious samples respectively from our dataset to conduct the following experiments. In training stage, we divide these 18,000 malware and 18,000 benign apps into three parts, 80\% of them are configured as training data, other 20\% are equally split into validating and testing data.}

\begin{table}\scriptsize
\caption{Used Dataset}
\label{tab:dataset}
\centering
\sen{\begin{tabular}{lccc}
\hline
{\bf Datasets} & {\bf Labels}  &{\bf Original Size} & {\bf Deduplicated Size} \\ \hline
Drebin  & Malware & 5,560   & 3,561   \\
Genome  & Malware & 1,260   & 1,009   \\
Contagio    & Malware & 360 & 198     \\
Pwnzen  & Malware & 1,830   & 572   \\
VirusShare  & Malware & 20,000  & 12,660  \\ \hline
{Total}  & Malware & 29,010  & 18,000  \\ \hline
{Total}  & Benign & 45,284  & 18,000  \\ \hline
\end{tabular}}
\end{table}

\paragraph{Setup} \label{experiment:feature_selection:updating:setup}
\updated{Because our previous work MobiDroid~\cite{feng2019mobidroid} applied three types of features (i.e., API calls, manifest properties, and opcode sequence), in this experiment, we determine to take API calls and manifest properties (i.e., 1,509 and 613), which our behavior-based feature updating method may benefit on, as the feature of MobiDroid to reveal the improvement on detection. Meanwhile, the updated version used in MobiTive has 2,290 API calls and 625 manifest properties. For each feature version, we apply three kinds of deep neural networks, which presented in \S\ref{approach:dl_model:training}, to investigate whether the feature updating method can improve the accuracy of our system.}

\paragraph{Results} \label{experiment:feature_selection:updating:results}
{In Table~\ref{tab:detection_feature_update}, the accuracy of updated feature version on CNN, LSTM and GRU is 95.11\%, 96.56\% and 96.75\%. Comparing to the previous results, there is around 1\%$\sim$5\% improvement after feature updating. 
Therefore, based on the result, we accept updating features summarised from potential malicious behaviors a part of our input feature \revised{set}.}

\begin{table}\scriptsize
\centering
\caption{Detection Results of Feature Updating}
\label{tab:detection_feature_update}
\sen{\begin{tabular}{lcccc}
\toprule
{\bf Feature Dictionary Size} & {\bf Networks} & {\bf Accuracy} & {\bf Precision} & {\bf Recall} \\ \midrule
1509 API + 613 Manifest    & CNN   & 90.03\%   & 91.00\%   & 89.26\%\\ \rowcolor{gray!30}
2290 API + 625 Mniafest    & CNN   & 95.11\%   & 95.06\%   & 95.16\%\\
1509 API + 613 Manifest    & LSTM  & 95.44\%   & 95.56\%  & 95.34\%\\ \rowcolor{gray!30}
2290 API + 625 Mniafest    & LSTM  & 96.56\%   & 96.72\%   & 96.40\%\\
1509 API + 613 Manifest    & GRU   & 95.56\%   & 95.22\%   & 95.86\%\\ \rowcolor{gray!30}
2290 API + 625 Mniafest    & GRU   & 96.75\%   & 96.78\%   & 96.72\%\\
\bottomrule
\end{tabular}}
\end{table}


\begin{table}\scriptsize
\centering
\caption{Detection Results of Feature Categories and Networks}
\label{tab:detection_categories_networks}
\sen{\begin{tabular}{lcccc}
\toprule
{\bf Categories} & {\bf Neural Networks} & {\bf Accuracy} & {\bf Precision} & {\bf Recall} \\ \midrule
Manifest & CNN   & 79.89\%   & 79.50\%   & 80.12\%\\
API Calls           & CNN   & 93.17\%   & 92.33\%   & 93.90\%\\ \hline
Two Types   & CNN   & 95.11\%   & 95.06\%   & 95.16\%\\
Two Types   & LSTM  & 96.56\%   & 96.72\%   & 96.40\%\\ \rowcolor{gray!30}
Two Types   & GRU   & 96.75\%   & 96.78\%   & 96.72\%\\ \hline
Two Types   & Stacked LSTM  & 96.64\%   & 96.83\%   & 96.46\%\\
Two Types   & Stacked GRU   & 96.67\%   & 97.00\%   & 96.36\%\\ \hline
Two Types   & Bidirectional LSTM  & 96.61\%   & 96.67\%   & 96.56\%\\ \rowcolor{gray!30}
Two Types   & Bidirectional GRU   & 96.78\%   & 97.00\%   & 96.57\%\\
\bottomrule
\end{tabular}}
\end{table}

\subsubsection{Accuracy evaluation of feature category selection and deep neural network selection} \label{experiment:feature_selection:categories_NN}
\updated{In this experiment, to find out the correlation between selected features and the effectiveness of different deep neural networks on detection accuracy, we first evaluate the effect of two newly-updated feature categories (Table~\ref{tab:selected_feature}) on detection accuracy separately. Second, we investigate the effect of computational architecture in different deep neural networks on detection accuracy.}

\paragraph{Dataset} \label{experiment:feature_selection:categories_NN:dataset}
\updated{The dataset configuration used in this experiment is same as \S\ref{experiment:feature_selection:updating:dataset} (Table~\ref{tab:dataset}).}

\paragraph{Setup} \label{experiment:feature_selection:categories_NN:setup}
\updated{To find out the correlation between the two selected features (i.e., manifest properties and API calls), we investigate their corresponding accuracy by accepting both single and combined feature categories as the input of a same neural network with a same training data configuration.
To determine the best deep neural network, we evaluate seven widely-used neural networks by using the combined two feature categories.}

\paragraph{Results} \label{experiment:feature_selection:categories_NN:results}
\updated{We demonstrate the results from the 2 aspects (feature category selection and network selection) as below.}

\noindent{(1) {Feature selection.}}
{As shown in Table~\ref{tab:detection_categories_networks}, the accuracy of the three CNN models is 79.89\%, 93.17\% and 95.11\%.
By comparing the accuracy of feature categories, we decide to use manifest properties and API calls together as an input bundle in our approach since the input with two feature types has the best result.}

\begin{figure}
    \includegraphics[width=0.34\textwidth]{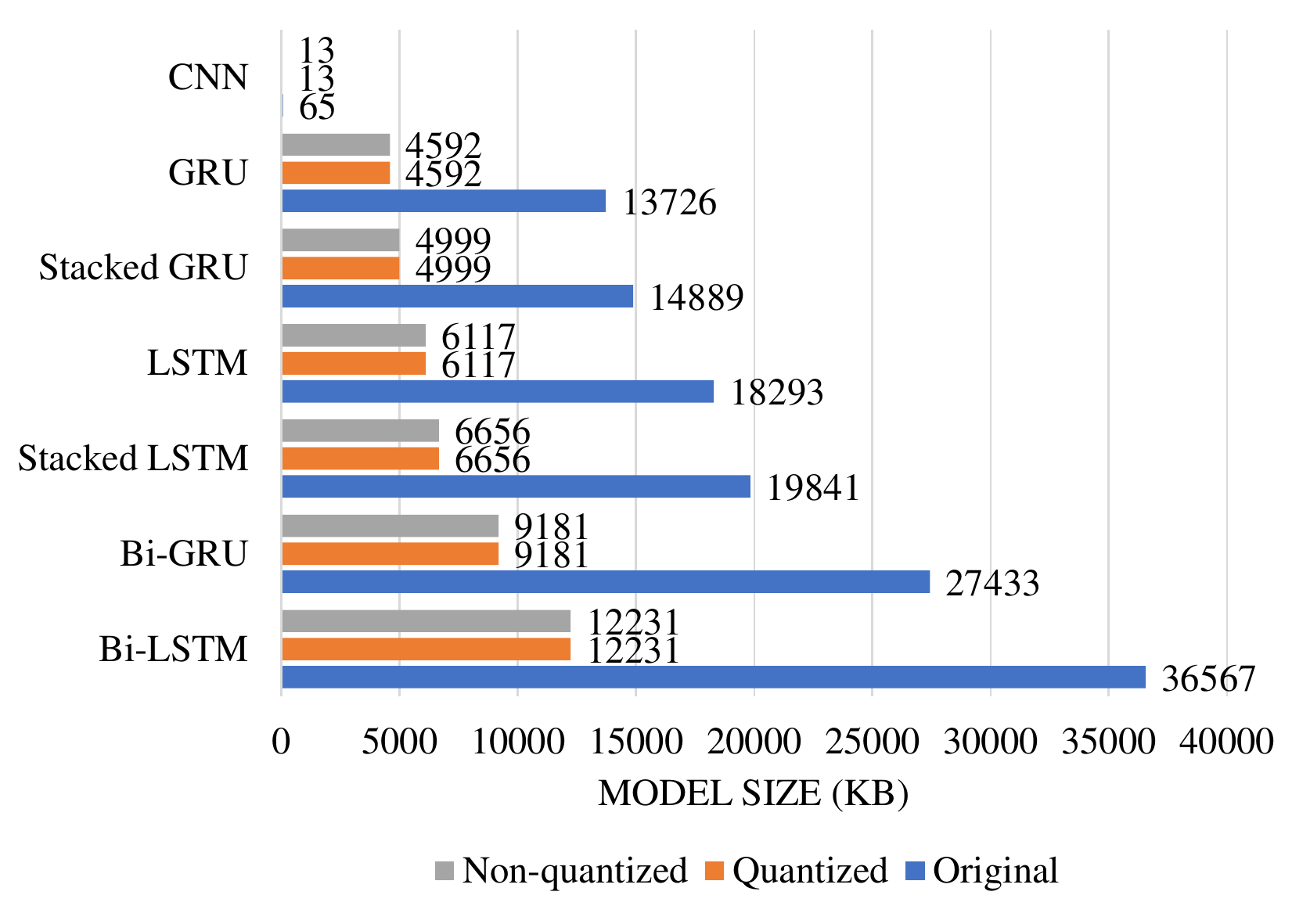}
    \centering
    
    \caption{\updated{Comparison of model size changes on migration and quantization}}
    \label{fig:model_size}
\end{figure}

\noindent{(2) {Network selection.}}
{In general, RNN models perform a better accuracy than CNN models.
A possible reason is that RNN has an internal state (memory), which can also take the correlation between the different feature positions into consideration. In the training stage, this internal state will make RNN be able to keep the highly potential related in a long-term and finally keeps the most corresponding feature positions. However, CNN considers every different feature position individually in training.
In terms of RNN models, GRU and bidirectional GRU achieve a similar accuracy (96.75\% vs. 96.78\%), which is better than other RNN models' accuracy.
They also have a better recall than precision (96.78\% vs. 96.72\% for GRU and 97.00\% vs. 96.57\% for Bidirectional GRU).
\updated{Besides, we also compare the size of original pre-trained model with the quantized and non-quantized models. In Fig.~\ref{fig:model_size}, we can find that the size of the original pre-trained model reduces 3 times on RNNs and 5 times on CNN by migrating it to the TensorFlow Lite model. In the experiment, whether the quantization configuration is enabled or not, the migrated model size for both RNN and CNN models will keep unchanged.}}

\subsubsection{{Comparison between the existing learning-based Android malware detection systems and MobiTive}} \label{experiment:feature_selection:comparison_other_work}
\updated{In this experiment, we evaluate our MobiTive together with several existing learning-based Android malware detection systems on both two directions, effectiveness and efficiency.}

\paragraph{Dataset} \label{experiment:feature_selection:comparison_other_work:dataset}
\updated{The dataset configuration used in this experiment is same as \S~\ref{experiment:feature_selection:updating:dataset} (Table~\ref{tab:dataset}).}

\paragraph{Setup} \label{experiment:feature_selection:comparison_other_work:setup}
\updated{We briefly compare our MobiTive with 4 open-source learning-based Android malware detection approaches, which applies different types of features (i.e., vector, sequence, and graph) as their inputs. 
There are three reasons to help us illustrating why we select these four approaches. By conducting a study on the corresponding literature which published in recent years, we first survey them on the feature types, and then select one representative work from each organized column. Further, by searching on the Github and sending emails to the authors, we obtain the source code of these four approaches and further evaluate them with our dataset to provide a more concrete comparison.
Since one of our basic concept in this paper is balancing the performance and accuracy to satisfy user's real usage, we not only evaluate the accuracy, but also compare the average feature extraction time for each approach on our dataset.}

\paragraph{Results} \label{experiment:feature_selection:comparison_other_work:results}
\updated{As shown in Table~\ref{tab:comparison_other_work}, comparing to McLaughlin et al.~\cite{mclaughlin2017deep}, MalDozer~\cite{karbab_maldozer_2018}, Apk2Vec~\cite{apk2vec}, there are obvious improvements on both the accuracy and feature extraction time cost on PC. Considering the approaches based on sequential features, the accuracy of MobiTive is higher than McLaughlin et al.~\cite{mclaughlin2017deep} and MalDozer (96.75\% vs. 94.79\% and 96.25\%), and the time cost of extracting feature is almost improved for 100 and 70 times than them (0.051 vs. 5.065 and 3.515 seconds). Considering our previous work, MobiDroid~\cite{feng2019mobidroid}, the accuracy of MobiTive is a little lower than MobiDroid (96.75\% vs. 96.87\%), however, with a tiny decrease at 0.12\% on the accuracy, the feature extraction time cost of MobiTive is almost 100 times shorter than MobiDroid (0.051 seconds vs. 5.892 seconds).}

\begin{table}\scriptsize
\centering
\caption{\updated{Comparison of MobiTive Against Existing Approaches}}
\label{tab:comparison_other_work}
\scalebox{0.9}{\begin{tabular}{lcccc}
\toprule
{\bf Feature Type} & \tabincell{c}{{\bf Extraction}\\{\bf Time (s)}} & \tabincell{c}{{\bf Classification}\\{\bf Method}} & {\bf Accuracy} & {\bf System} \\ \midrule
Opcode Seq                                              & 5.065     & \tabincell{c}{{Deep}\\ {Learning(CNN)}}        & 94.79\%  & \tabincell{c}{{McLaughlin}\\{et al.}}~\cite{mclaughlin2017deep}\\ \rowcolor{gray!30}
API Call Seq                                            & 3.515     & \tabincell{c}{{Deep}\\ {Learning(CNN)}}        & 96.25\%  & MalDozer~\cite{karbab_maldozer_2018}\\
iCFG                                                    & 198.920   & \tabincell{c}{{Representation}\\ {Learning}}   & 90.98\%  & Apk2Vec~\cite{apk2vec}\\ \rowcolor{gray!30}
\tabincell{l}{{Manifest}\\ {API Call} \\{Opcode Seq}}   & 5.892     & \tabincell{c}{{Deep}\\ {Learning(CNN)}}        & 96.87\%  & MobiDroid~\cite{feng2019mobidroid}\\
\tabincell{l}{{Manifest}\\ {API Call}}                  & 0.051     & \tabincell{c}{{Deep}\\ {Learning(GRU)}}        & 96.75\%  & MobiTive\\
\bottomrule
\end{tabular}}
\end{table}

\subsubsection{{Accuracy evaluation of multi-class classification}} \label{experiment:feature_selection:multi_class}
\updated{In this experiment, we evaluate the effectiveness of our MobiTive on predicting malware in different virus types such as Spyware and Trojan.}

\paragraph{Dataset} \label{experiment:feature_selection:multi_class:dataset}
\updated{We collect 70,130 Android applications from VirusShare as shown in Table~\ref{tab:family_acc} and classify them with VirusTotal~\cite{virustotal}, which is an online detection platform, to retrieve their types as our ground truth in multi-class classification. Finally, we have 21 virus-labels of these Android malware, which locate in 5 types (i.e., Adware, Spyware, Riskware, Trojan, and File-Infector). In training stage, we also accept the same data split in binary classification (i.e., 80\%, 10\% and 10\%) as the train/validate/test data split portion.}


\paragraph{Setup} \label{experiment:feature_selection:multi_class:setup}
\updated{To evaluate the effectiveness on classifying malware into different virus types, we train a multi-class malware classifier on our collected dataset in Table~\ref{tab:family_acc} with the determined deep neural network (i.e., GRU).}

\paragraph{Results} \label{experiment:feature_selection:multi_class:results}
\updated{As shown in Table~\ref{tab:family_acc}, we reach an overall accuracy at {94.45\%} in multi-class malware classification task. Considering those families respectively, on each of the two Riskware families, our MobiTive performs a perfect prediction, which has an accuracy at 100.00\%. The detection accuracy of Malware families, cnzz and commplat in File-Infector type, also reach 100.00\%, and the other family, domob, has an accuracy at 93.33\%. Each of the malware types located in Trojan achieves an accuracy above 95\% (i.e., 95.92\%, 97.28\%, 97.47\%, and 100.00\%). For the only spyware, the accuracy reaches 99.74\% among 391 test malware applications. For Adware families, with our selected features, MobiTive achieves an accuracy above 95.00\% in detecting anydown, baiduprotect, feiwo, fictus, gappusin, leadbolt (i.e., 100.00\%, 98.35\%, 95.86\%, 96.53\%, 96.16\%, and 96.80\%), however, it fails to provide a dependable prediction result on admogo, adwo, dowgin, kuguo, kyview (i.e., 84.38\%, 86.08\%, 78.88\%, 88.83\%, 79.72\%).}

\begin{table}\scriptsize
\caption{\updated{Detection Result with Multi-Class Dataset}}
\label{tab:family_acc}
\centering
\begin{tabular}{lcccc}
\toprule
{\bf Name}    &   {\bf Type}    &   {\bf \#Total Data}    &   {\bf \#Test Data}  &   {\bf Accuracy}   \\ \midrule
admogo          & Adware    &   1,918   &   192 &    84.38\%\\\rowcolor{gray!30}
adwo            & Adware    &   3,882   &   388 &    86.08\%\\
airpush         & Trojan    &   6,368   &   637 &    95.92\%\\\rowcolor{gray!30}
anydown         & Adware    &   1,343   &   134 &    100.00\%\\
baiduprotect    & Adware    &   3,026   &   303 &    98.35\%\\\rowcolor{gray!30}
cnzz            & File-Infector & 970   &   97  & 100.00\%\\
commplat        & File-Infector & 1,442 &   144 & 100.00\%\\\rowcolor{gray!30}
domob           & File-Infector & 5,696 &   570 & 93.33\%\\
dowgin          & Adware    &   3,223   &   322 &   78.88\%\\\rowcolor{gray!30}
feiwo           & Adware    &   1,694   &   169 &   95.86\%\\
fictus          & Adware    &   1,435   &   144 &   96.53\%\\\rowcolor{gray!30}
gappusin        & Adware    &   9,378   &   938 &   96.16\%\\
igexin          & Spyware   &   3,911   &   391 &   99.74\%\\\rowcolor{gray!30}
jiagu           & Riskware  &   2,662   &   266 &   100.00\%\\
kuguo           & Adware    &   4,031   &   403 &   88.83\%\\\rowcolor{gray!30}
kyview          & Adware    &   1,433   &   143 &   79.72\%\\
leadbolt        & Adware    &   5,929   &   593 &   96.80\%\\\rowcolor{gray!30}
mecor           & Riskware  &   833     &   83  &   100.00\%\\
plankton        & Trojan    &   2,571   &   257 &   97.28\%\\\rowcolor{gray!30}
revmob          & Trojan    &   7,517   &   752 &   97.47\%\\
scamapp         & Trojan    &   868     &   87  &   100.00\%\\\midrule
Overall         &       &   70,130  &   7,013  &   94.45\%\\\bottomrule
\end{tabular}
\end{table}

	\noindent{\textbf{Remark.} To face the high latency during feature preparation, we find extracting API calls and manifest properties from unzipped Dalvik binary and binary manifest file will cost less than 1 second.
	To validate the effect of our newly supplemented features, we find the RNNs have an improvement at over 1\% on the accuracy, and the accuracy of CNN increased by 5\%.
	Meanwhile, by comparing the result on different feature categories and deep neural networks, we find that (1) two feature types combined input has a much better result than single feature type; (2) on average, the RNN models have a better result than CNN. GRU models have a better accuracy than the LSTM models on our dataset. \updated{Moreover, by comparing 4 existing approaches with MobiTive, it achieves a better performance with a considerable detection accuracy.
	To validate the effect on multi-class classification, we find that MobiTive can efficiently handle most malware families (i.e., 17/21 obtain an accuracy larger than 95\%).}}

\subsection{Effectiveness Evaluation of MobiTive on Mobile Devices} \label{experiment:mobile}

\subsubsection{Performance evaluation of feature preparation on real devices} \label{experiment:mobile:preparation}
\updated{We evaluate the performance of feature preparation on real mobile devices in this experiment. The time cost of feature preparation step includes unzipping time and feature extraction time.}

\paragraph{Dataset} \label{experiment:mobile:preparation:dataset}
\updated{The dataset configuration used in this experiment is same as \S\ref{experiment:feature_selection:raw_data_feature_extraction:dataset}.}

\paragraph{Setup} \label{experiment:mobile:preparation:setup}
\updated{We first evaluate the performance of raw data extraction by {investigating} the time cost of unzipping the applications with 6 different sizes on 6 different real Android devices. Second, with the extracted raw data (i.e., binary manifest file and Dalvik executable file), we further evaluate the performance of feature extraction by {investigating} the time cost of extracting the features from raw data on the devices.}

\paragraph{Results} \label{experiment:mobile:preparation:results}
\updated{We introduce the results from 2 aspects (unzipping time and feature extraction time) as below.}

\begin{figure}
    \includegraphics[width=0.45\textwidth]{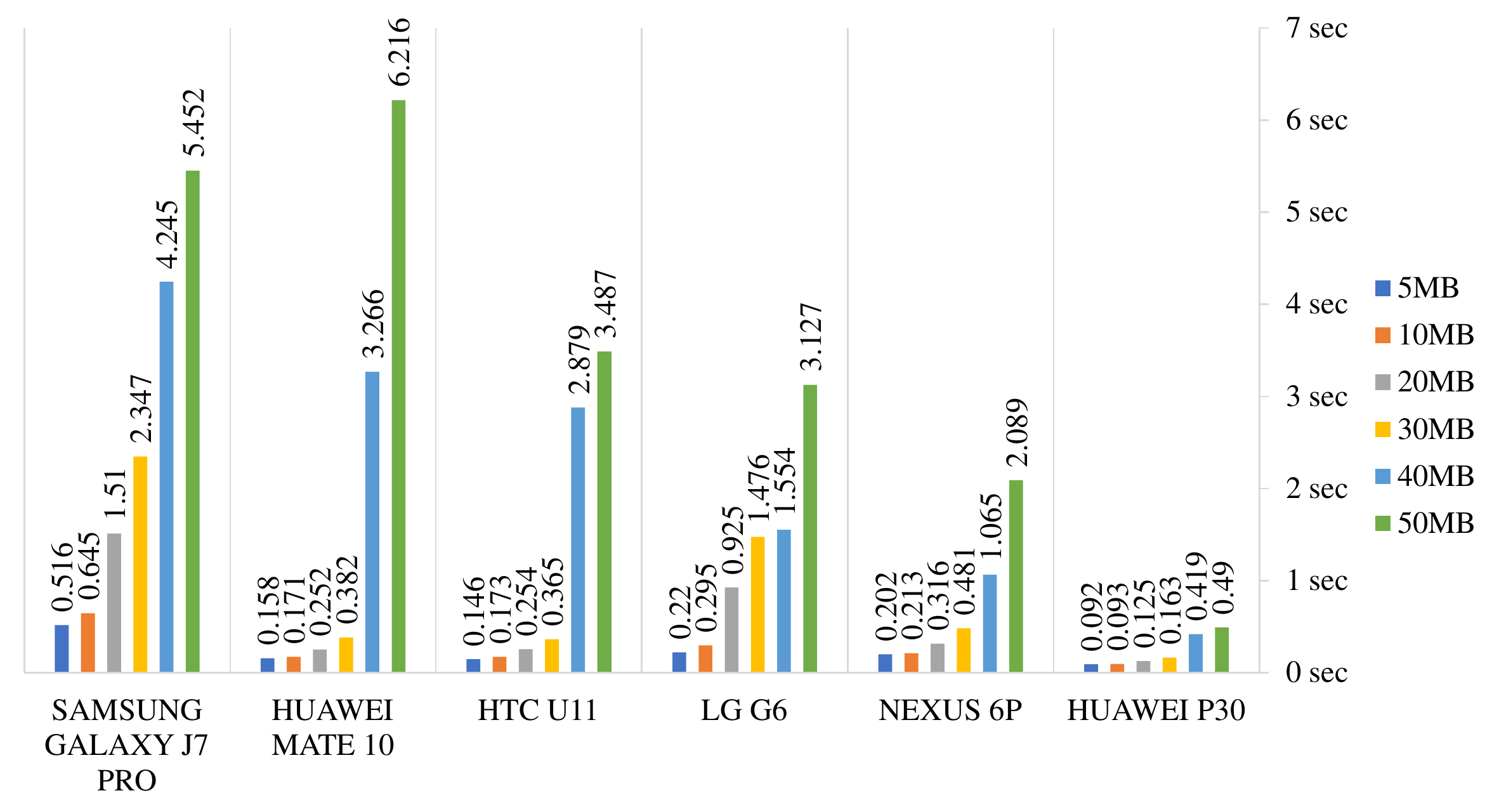}
    \centering
    \caption{\sen{Unzipping time on different mobile devices}
    }
    \label{fig:time_unzip}
\end{figure}

\noindent{(1) \textit{Unzipping time evaluation on real devices.}}
{Fig.~\ref{fig:time_unzip} shows the average unzipping time of 50MB apps on common specification devices (Huawei Mate 10, HTC U11, Nexus 6P, LG G6) locates between 1.023 and 2.586 seconds.
For 5MB apps, it locates between 0.119 and 0.264s.
For the performance of low-profile device (Samsung Galaxy J7 Pro), the unzipping time of 5MB and 50MB apps are 0.261 and 3.918s.
Considering flagship device (Huawei P30), they are limited to less than 0.6 second. For 5MB apps, it only takes 0.046s.}

\begin{figure}
    \includegraphics[width=0.45\textwidth]{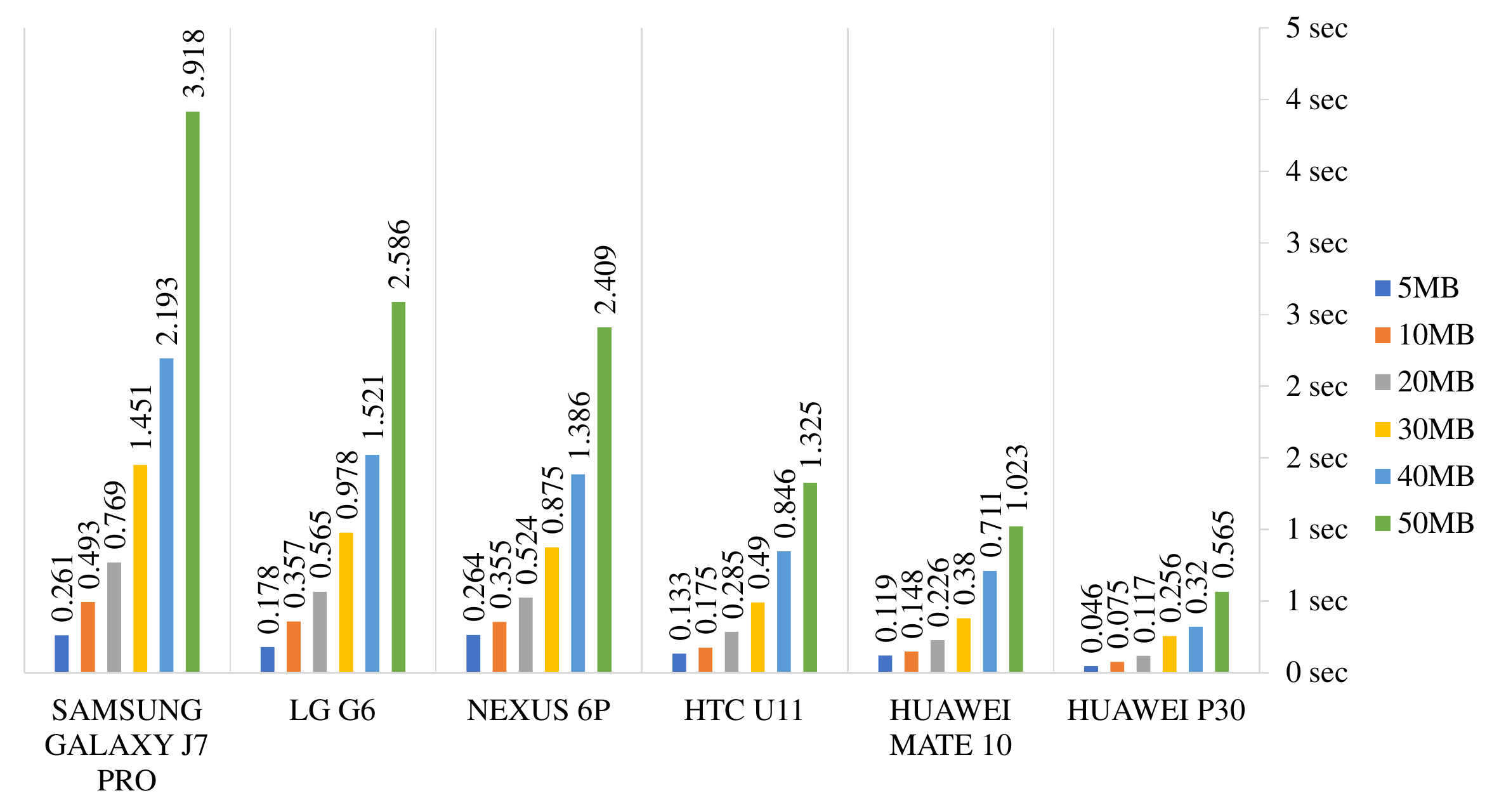}
    \centering
    \caption{\sen{Extraction time on different mobile devices}
    }
    \label{fig:time_extract}
\end{figure}

\noindent{(2) \textit{Feature extraction time evaluation on real devices.}}
{Apart from the performance evaluation of unzipping time, we further evaluate the feature extraction time.
To extract the API calls, we package our API call parser used on server side into a jar with the help of {JitPack}. 
The API call parser is used to extract API calls from binary code.
Since the XML decoder (axmldec) used on sever is implemented in C++, we apply {AXML} as a lib to extract the manifest properties, which is more suitable on the mobile side. Fig.~\ref{fig:time_extract} shows the average feature extraction time of 50MB apps on common specification devices locates between 2.089 and 6.216 seconds.
For 5MB apps, it locates between 0.146 and 0.22 seconds.
On low-profile device (Samsung Galaxy J7 Pro), the extraction time of 5MB and 50MB apps are 0.516 and 5.452 seconds.
Considering flagship device (Huawei P30), they are limited to less than 0.49 second. For 5MB apps, it only takes 0.092 second, which is very fast.}

\subsubsection{Performance evaluation of RNN models on real device} \label{experiment:mobile:performance_RNN_mobile}

\updated{Besides analyzing the performance of unzipping and extracting features, in this experiment, we further evaluate the {efficiency of prediction with different RNN models} on real devices.}

\paragraph{Dataset} \label{experiment:mobile:performance_RNN_mobile:dataset}
\updated{To make sure that the test accuracy is comparable to the results obtained on server, the testing data used in this mobile-end experiment is same to the testing data generated by the data split function mentioned in \S\ref{experiment:feature_selection:updating:dataset} (Table~\ref{tab:dataset}), including 1,800 malware and 1,800 benign samples respectively. To get rid of the influence of feature preparation phase in this performance evaluation {against} RNN models, we directly use {a set of feature vectors extracted from testing data as the input.}}

\paragraph{Setup}\label{experiment:mobile:performance_RNN_mobile:setup}
\updated{We first convert and migrate the RNN models obtained in \S\ref{experiment:feature_selection:categories_NN} (i.e., simple RNN LSTM/GRU, stacked LSTM/GRU, and bidirectiontal LSTM/GRU) to TensorFlow Lite models and further deploy them on real device (e.g., Huawei P30). Secondly, to provide an insight for both the prediction accuracy and performance, we investigate the prediction time for each model with our dataset and organize the result together with the accuracy obtained in \S\ref{experiment:feature_selection:categories_NN} (Table~\ref{tab:detection_categories_networks}).}

\begin{figure}
    \includegraphics[width=0.4\textwidth]{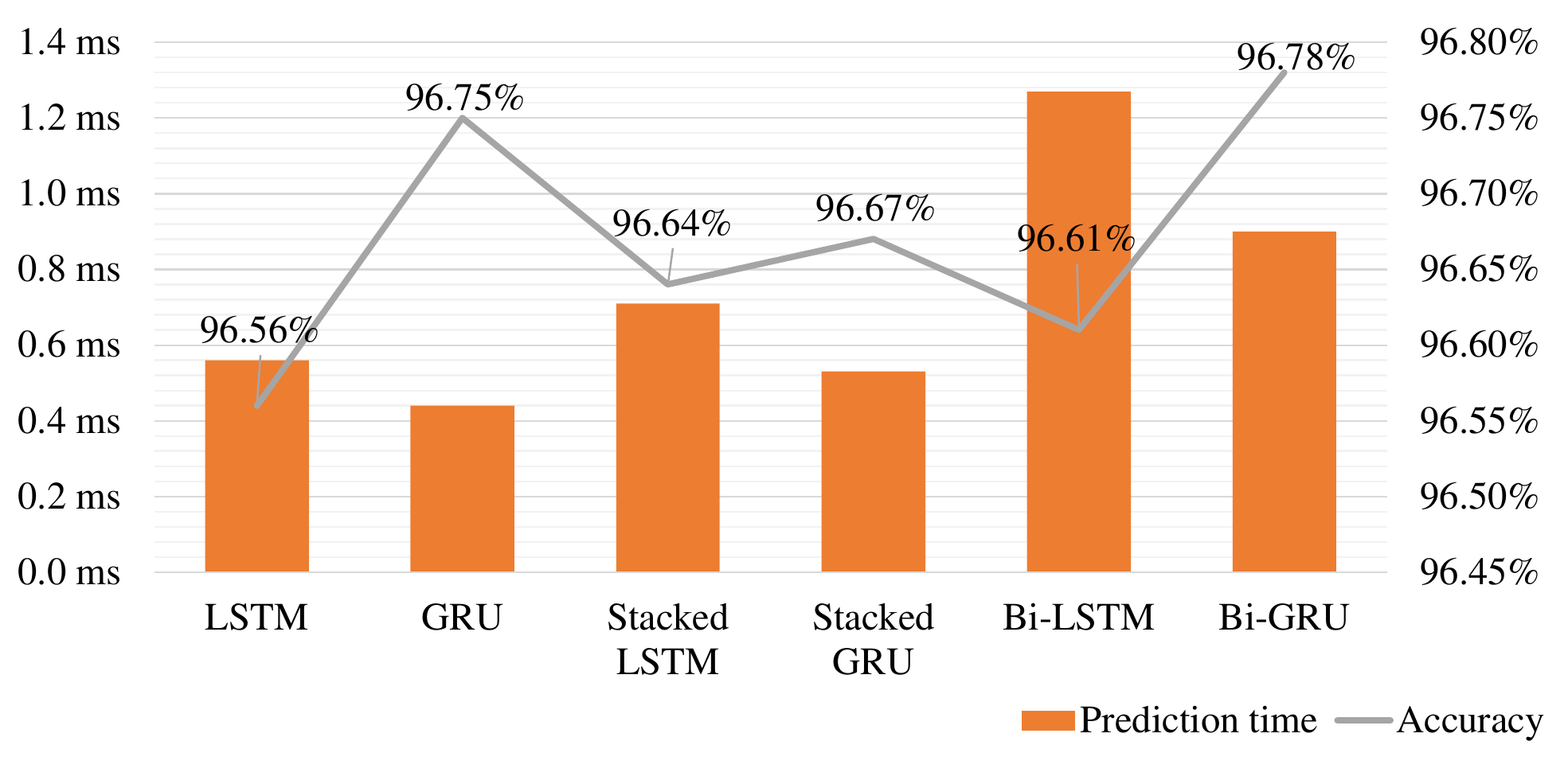}
    \centering
    \caption{\sen{The accuracy and prediction time of RNN models on Huawei P30}}
    \label{fig:acc_pred_time_RNN}
\end{figure}

\paragraph{Results} \label{experiment:mobile:performance_RNN_mobile:results}
{As shown in Fig.~\ref{fig:acc_pred_time_RNN}, by comparing the prediction time of different RNN models, which presented in the histogram, we can see that the pre-trained model with GRU has the best performance than any others. 
Meanwhile, from the grey accuracy line in this figure, we can see it has the second highest accuracy among them, which only has a small difference comparing to the accuracy of bidirectional GRU (96.75\% vs. 96.78\%).
Considering all situations, we select GRU model to evaluate the performance and accuracy of our approach on the real mobile devices.
}

\subsubsection{Accuracy and prediction time on different real devices} \label{experiment:mobile:acc_time_mobile}
\updated{In this experiment, we evaluate the effectiveness of MobiTive on real Android mobile devices by conducting a comparison experiment on test accuracy and the total prediction time on real devices.}

\paragraph{Dataset} \label{experiment:mobile:acc_time_mobile:dataset}
\updated{To make sure that the test accuracy is comparable to the results obtained on server, the testing data used in this mobile-end experiment is same to the testing data generated by the data split function mentioned in \S\ref{experiment:feature_selection:updating:dataset} (Table~\ref{tab:dataset}), including 1,800 malware and 1,800 benign samples.}

\paragraph{Setup} \label{experiment:mobile:acc_time_mobile:setup}
\updated{We first convert and migrate the GRU model obtained in \S\ref{experiment:feature_selection:categories_NN} to quantized/non-qunatized TensorFlow Lite models and deploy them on real devices. Second, we record the average prediction time and detection accuracy by testing the quantized/non-quantized GRU models. Note that, to provide an insight for the performance of each processing phase, we record the time cost by 3 parts (i.e., raw data unzipping, feature extraction, and prediction).}

\paragraph{Results} \label{experiment:mobile:acc_time_mobile:results}
\updated{With the obtained GRU model in \S\ref{experiment:feature_selection:categories_NN} (accuracy: 96.75\%), in Table~\ref{tab:accuracy_performance_real_device}, by comparing the accuracy of non-quantized and quantized models, we find that the accuracy of quantized model will almost equal to the non-quantized model for RNN (GRU).}
However, by comparing the prediction time of them, it shows that the performance of predicting with quantized model is a little better than non-quantized model.
In this experiment, the result shows that the difference of prediction time, which brought by the quantization technique, is less than 0.01 microseconds.
As a result of the current inadequate support for the operators in Tensorflow Lite, the structure of current applied deep neural networks are relatively simple.
However, with a more complicated neural network, the quantization technique will definitely provide a performance boost during the prediction phase~\cite{guo2019empirical}.

{By calculating the unzipping, analyzing, and prediction time together, the time is always acceptable for mobile users (i.e., less than 3 seconds on average, less than 1 second in best practice).
By comparing the specifications of these devices used in our experiment with the summarized devices' specifications (details on our website~\cite{mobitive}), we find that the performance benchmark result of most newly released devices are better than the common devices selected in our experiments.
Thus, we can claim that the current mobile phones can support our off-line prediction system smoothly.}

\begin{table}\scriptsize
\centering
\caption{\updated{Run-Time Performance Comparison of MobiTive Against Dynamic Android Analysis Tool}}
\label{tab:comparison_dynamic_work}
\scalebox{0.95}{\begin{tabular}{ccccc}
\toprule
\tabincell{c}{{\bf CPU}\\ {\bf Usage}} & \tabincell{c}{{\bf Memory}\\{\bf Usage (MB)}} & \tabincell{c}{{\bf Energy}\\{\bf Usage}} & \tabincell{c}{{\bf Execution}\\ {\bf Time (s)}} & {\bf System} \\ \midrule
$P_{40\%}$            & 60            & $Q_{M}\times0.46$        &   0.46        & MobiTive\\ \hline
$P_{10\%}\times n$ & $130\times n$ & $Q_{L}\times t\times n$   &   $\infty$    &   Inspeckage~\cite{Inspeckage}\\ \bottomrule
\end{tabular}}
\begin{tablenotes}
\item {1.} $t$ is the execution time to finish one detection on an app.
\item {2.} $n$ is the total number of monitored apps running in foreground or background. 
\item {3.} $P$ is the run-time CPU power usage in one detection.
\item {4.} $Q$ is the run-time energy cost in one detection, $Q_{M}$ and $Q_{L}$ represent the energy cost at level medium and light (defined by Android Profiler~\cite{Android-profiler}).
\end{tablenotes}
\end{table}

\noindent{\textit{Composition of overall time analysis.}}
{Comparing the feature preparation and prediction time in common cases in Table~\ref{tab:accuracy_performance_real_device}.
we can see that the detection time only cost less than 1\% among the total time on common spec devices.
Thus, reducing the time cost in feature preparation will bring a considerable performance improvement for our detection system.
It is also a strong motivation for us.
Additionally, as a result of the installation mechanism, the downloaded Android APK will be always unzipped by the Android operation system.
Thus, there will be a same step between our approach and the installing procedures, which is extracting the same raw data from the target APK.
If we can deploy our approach on the Android operation system framework directly, the time cost of unzipping step in our approach will be saved. It is a new research point of this field in the future.}

\begin{table*}\scriptsize
\centering
\caption{\sen{Accuracy and performance of MobiTive on real mobile devices}}
\label{tab:accuracy_performance_real_device}
\begin{tabular}{c|c|c|c|c|c|c|c}
\hline
{\bf Devices}    &   \tabincell{c}{{\bf Released}\\{\bf Year}}   & \tabincell{c}{{\bf Unzipping}\\{\bf Time (s)}}   &   \tabincell{c}{{\bf Extraction}\\{\bf Time (s)}}   &    {\bf Quantized}    &   {\bf Accuracy}   &   \tabincell{c}{{\bf Prediction}\\{\bf Time (ms)}}   &   \tabincell{c}{{\bf Total}\\{\bf Time (s)}} \\ \hline
\multirow{2}{*}{Nexus 6P}               &   \multirow{2}{*}{Sep 2015}   &   \multirow{2}{*}{0.97}   &   \multirow{2}{*}{0.73}   &   No  &   96.75\% &   1.14                        &   1.70    \\ \cline{5-8} 
                                        &                               &                           &                           &   Yes &   96.75\% &   1.14                        &   1.70    \\  \hline
\multirow{2}{*}{LG G6}                  &   \multirow{2}{*}{Apr 2017}   &   \multirow{2}{*}{1.03}   &   \multirow{2}{*}{1.27}   &   No  &   96.75\% &   0.74                        &   2.30    \\ \cline{5-8} 
                                        &                               &                           &                           &   Yes &   96.75\% &   \cellcolor{grayone} 0.73    &   2.30    \\ \hline
\multirow{2}{*}{Samsung Galaxy J7 Pro}  &   \multirow{2}{*}{Jun 2017}   &   \multirow{2}{*}{1.51}   &   \multirow{2}{*}{2.45}   &   No  &   96.75\% &   1.67                        &   3.96    \\ \cline{5-8} 
                                        &                               &                           &                           &   Yes &   96.75\% &   \cellcolor{grayone} 1.66    &  \cellcolor{grayone}  3.96    \\ \hline
\multirow{2}{*}{HTC U11}                &   \multirow{2}{*}{Jun 2017}   &   \multirow{2}{*}{0.54}   &   \multirow{2}{*}{1.22}   &   No  &   96.75\% &   1.10                        &   1.76    \\ \cline{5-8} 
                                        &                               &                           &                           &   Yes &   96.75\% &   1.10                        &   1.76    \\ \hline
\multirow{2}{*}{Huawei Mate 10}         &   \multirow{2}{*}{Feb 2018}   &   \multirow{2}{*}{0.44}   &   \multirow{2}{*}{1.74}   &   No  &   96.75\% &   1.10                        &   2.18    \\ \cline{5-8} 
                                        &                               &                           &                           &   Yes &   96.75\% &   \cellcolor{grayone} 1.09    &   2.18    \\ \hline
\multirow{2}{*}{Huawei P30}             &   \multirow{2}{*}{Mar 2019}   &   \multirow{2}{*}{0.23}   &   \multirow{2}{*}{0.23}   &   No  &   96.75\% &   0.56                        &   0.46 \\ \cline{5-8} 
                                        &                               &                           &                           &   Yes &   96.75\% &   0.55                        &  \cellcolor{grayone} 0.46    \\  \hline
\end{tabular}
\end{table*}

\subsubsection{{Performance comparison between MobiTive and dynamic run-time detection}} \label{experiment:mobile:compare_dynamic}
\updated{In this experiment, we evaluate and discuss the efficiency of our MobiTive against other tool, which based on dynamic behavior analysis.}

\paragraph{Setup} \label{experiment:mobile:compare_dynamic:setup}
\updated{To evaluate the run-time performance of our MobiTive against other tool, which applied dynamic behavior analysis as the baseline technique, we select Inspeckage~\cite{Inspeckage}, which is a state-of-art tool developed to offer dynamic analysis for Android applications, as the target system in this experiment. We investigate the run-time performance cost of both Inspeckage and MobiTive on three aspects (i.e., CPU, memory, energy) with the help of {Android Profiler}~\cite{Android-profiler}.}

\paragraph{Results} \label{experiment:mobile:compare_dynamic:results}
\updated{In Table~\ref{tab:comparison_dynamic_work}, the result shows that if detecting a \textbf{single application} successfully in the \textbf{same limited time period}, the CPU usage and energy consumption of Inspeckage look better than MobiTive (i.e., CPU: $P_{40\%}$ vs. $P_{10\%}$ and $Q_M$ vs. $Q_L$), but the average memory usage is 70MB larger than ours (i.e., 60 vs. 130MB). More importantly, the differences of Inspeckage and MobiTive on their basic mechanisms cause that more factors need to be involved in this evaluation. First, the protection of MobiTive is only provided before the installation of target application. In other words, every application will only trigger MobiTive's detection once in each installation and cost 0.46s on average. Considering the protection of Inspeckage is provided by monitoring and analyzing the behaviors of applications in real time, namely, it will be assigned to every application, no matter running in foreground or background. Meanwhile, unlike MobiTive can be slept after every detection, the Inspeckage has to be kept running all the time if any applications are running. Thus, the CPU usage, memory usage and energy usage of Inspeckage will turn to be $P_{10\%}\times n$, 130$\times n$ and $Q_L\times t\times n$, where $t$ is the execution time to finish one detection on an application and $n$ is the total number of protected applications running in foreground/background. In conclusion, in common scenario of Android devices' usage, malware detection tool using dynamic analysis, like the Inspeckage, will definitely lead to a higher performance cost than MobiTive.}

    \noindent{\textbf{Remark.}} To determine the feature preparation and prediction performance on real devices, we find (1) the feature preparation time is less than 4 seconds on average; (2) the prediction time of RNN models is less than 2ms on average. GRU costs 0.44ms with the best performance.
    Meanwhile, by comparing the result on six mobile devices, we find MobiTive costs (1) less than 3 seconds on common devices, (2) less than 0.5 second on flagship device.
    \updated{Meanwhile, by comparing the run-time performance of MobiTive and Inspeckage, we find that MobiTive can serve user with a much more efficient experience as {an} on-device protection system.}

\subsection{Analysis of Hardware Performance Evolution Trend of Android Mobile Devices} \label{experiment:performance_trend}
\sen{In this section, we conduct a study from three different aspects to investigate the hardware performance evolution trend of Android mobile devices.}

\begin{figure}
    \includegraphics[width=0.49\textwidth]{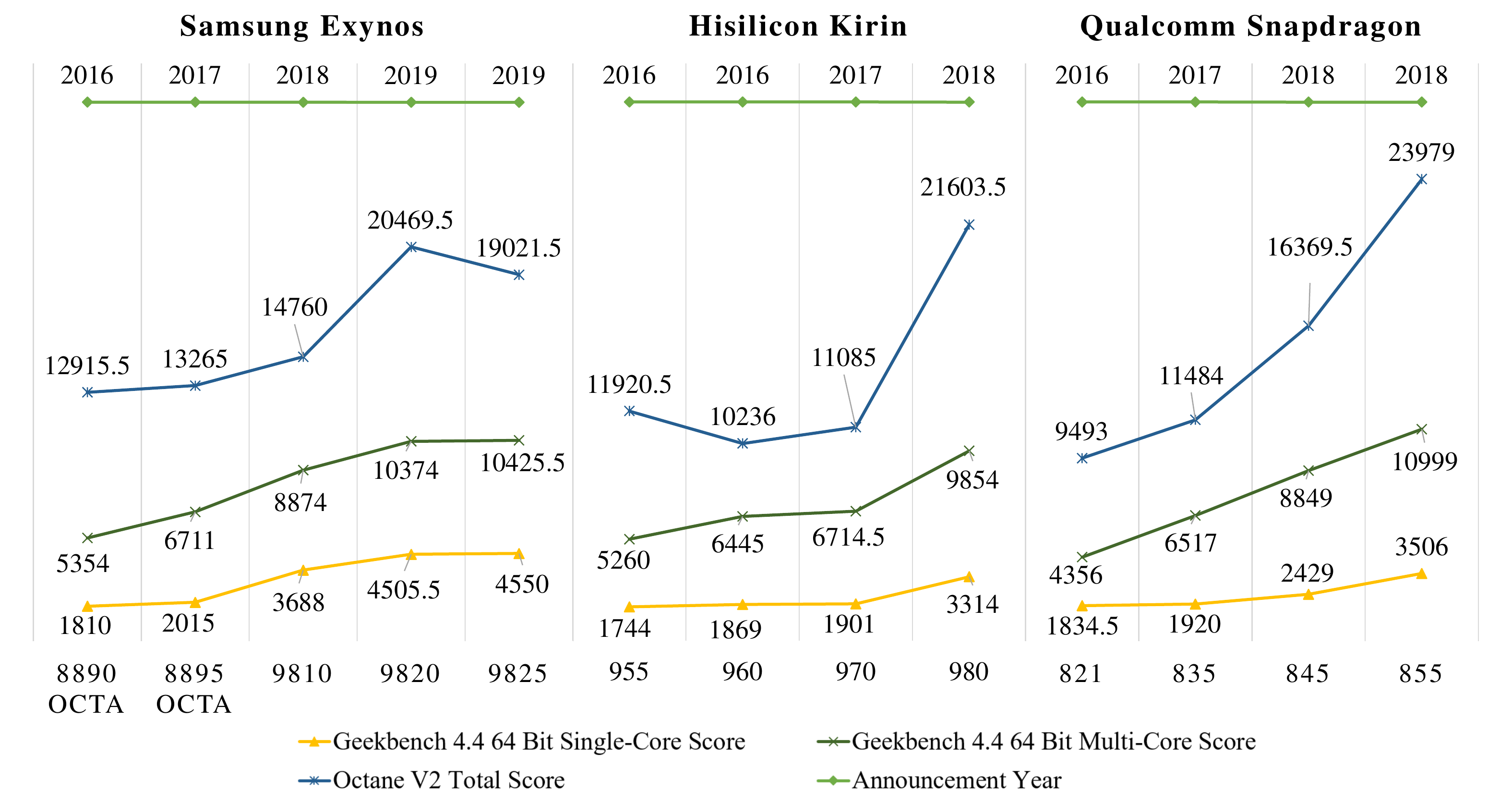}
    \centering
    \vspace{-5mm}
    \caption{Benchmark score comparison among different chipsets from Exynos, Kirin, and Snapdragon}
    \label{fig:benchmark_chipset}
\end{figure}

\begin{figure}
    \includegraphics[width=0.45\textwidth]{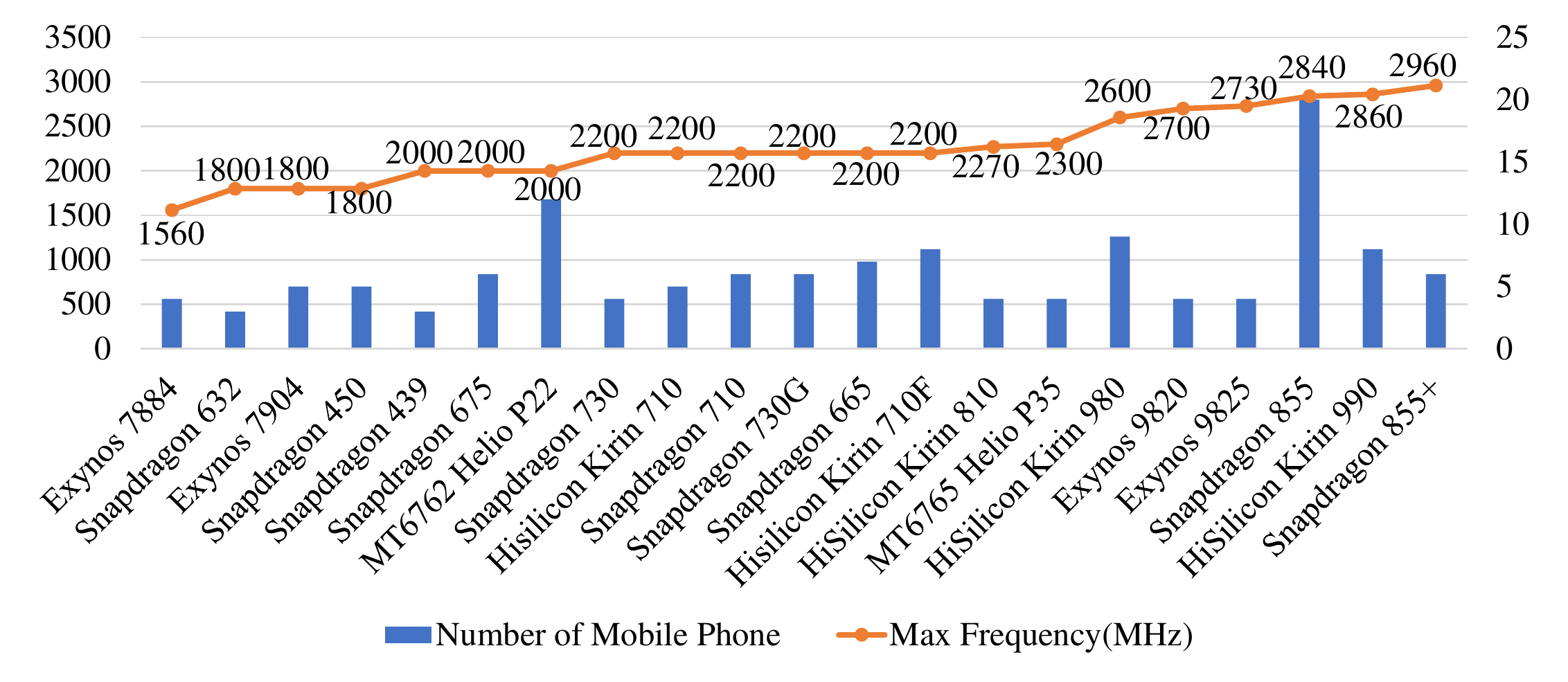}
    \centering
    \vspace{-2mm}
    \caption{\sen{Top 21 chipsets assembled in Android mobile phones released in 2019}}
    \label{fig:top_21_chipset}
\end{figure}

\begin{figure}[t]
    \includegraphics[width=0.2\textwidth]{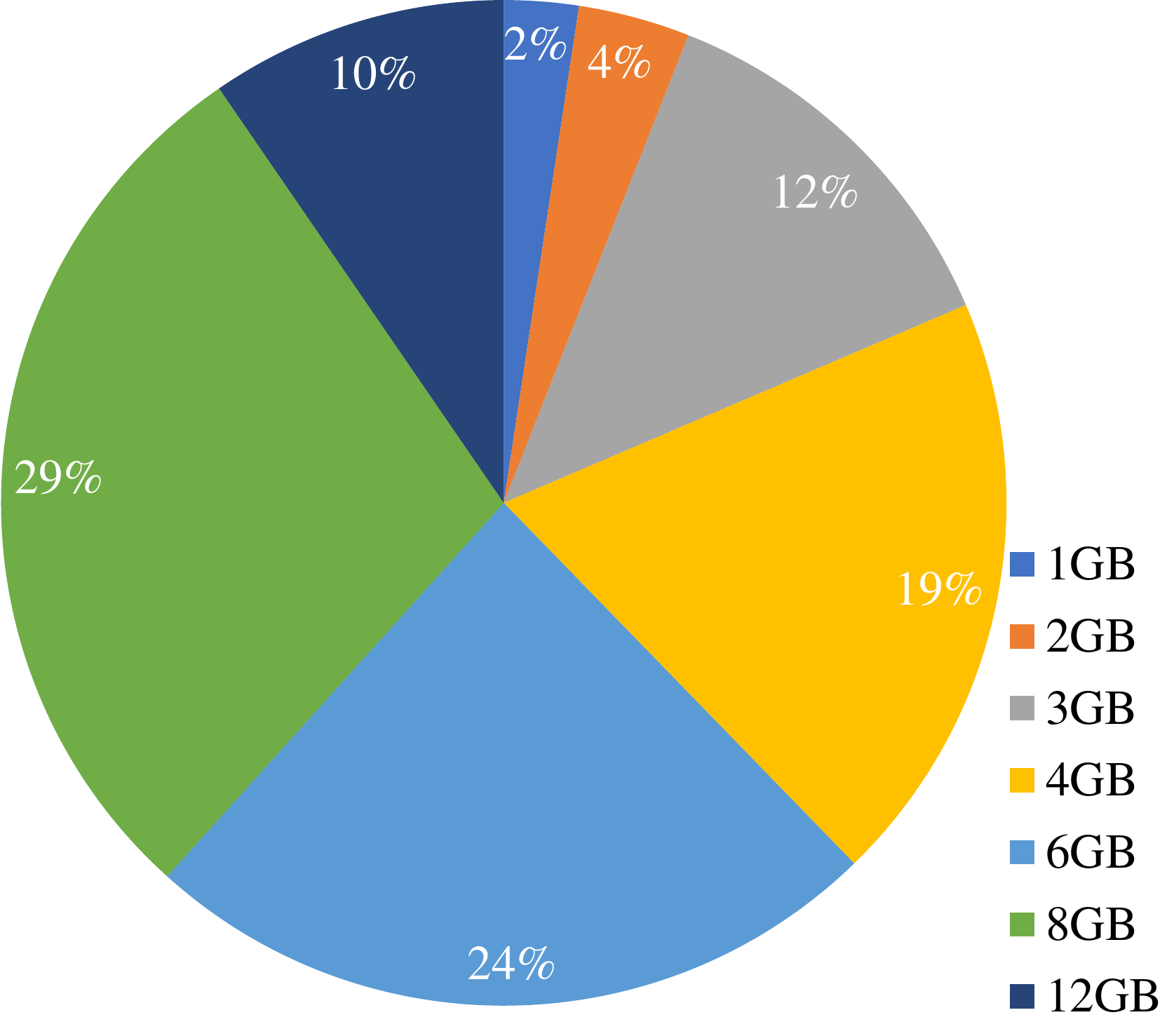}
    \centering
    \caption{\sen{RAM size of Android mobile devices released in 2019}}
    \label{fig:ram}
\end{figure}

\sen{To provide insights into the current and future usability of MobiTive, we study 45 widely-used chipsets, which released between 2016 and 2019.
They are collected from three well-known brands, Exynos, Kirin, and Snapdragon
We select 4--5 chipsets from each brand and compare them with 3 different kinds of benchmark test scores (i.e., Greekbench 4.4 64 Bit Single-Core score, Greekbench 4.4 64 Bit Multi-Core score, and Octane V2 total score).
As shown in Fig~\ref{fig:benchmark_chipset}, we present the results along time line to reveal the fast evolution trend of the chipsets.
From the polylines, which refer to the different score results, we can see that the performance of the chipsets has doubled during the past 5 years.
We detail the full specifications of chipsets on our website~\cite{mobitive}.
{Besides the analysis of chipsets, we also investigate the clock freqency and RAM size on 167 Android mobile devices, which is released in 2019, and present them on our website~\cite{mobitive}.
As shown in Fig.~\ref{fig:top_21_chipset}, we can see that the current frequency of new released Android devices are mostly located in 2000$\sim$2500MHz and 2500$\sim$3000MHz, which refers to common devices and flagship devices respectively.
}
As shown in Fig.~\ref{fig:ram}, we can see that the current RAM sizes of new released Android devices are mostly larger than 3GB.
The mainstream RAM sizes are 4GB, 6GB, and 8GB, which have a proportion around 72\% among the whole specification data.
By investigating the hardware performance evaluation results of real devices on chipsets and RAM with the six devices used in our experiments, we can tell that the most of the current Android mobile phones can support MobiTive smoothly and achieve a responsive detection.
}

	\noindent{\textbf{Remark.} By collecting and analyzing the specifications of chipsets and devices, we find the evolution trend of chipsets will provide a better performance for MobiTive in the future.
	Meanwhile, the study on device specifications released in 2019 shows that most new devices will have a performance not worse than our 4 selected common devices.}

\section{Limitations and Discussions} \label{limitation}

\noindent \textbf{Feature selection.} \label{limitation:feature}
\sen{As a result of the performance requirement of MobiTive, the limited selected feature categories (i.e., manifest properties and API calls) surely will not cause large overhead when it is working on an Android device. However, the limited two feature types will also provide limited information from the Android malware. If there will be a new malware family, whose malicious behaviors can not be represented by our selected feature types, the MobiTive may not be able to detect them. In the future, we aim to add more effective feature types with low-performance costs as well.}

Meanwhile, it is very important to detect new malware families in practice. Actually, neither dynamic nor static methods can fully guarantee the validity of protection against the new malware samples. A possible solution for detecting more new malware is that combining two types of methods. In the future, we will try to improve the ability to detect new malware by designing a new adaptive method, which is also an open question for this community.

\noindent \textbf{New malware family detection.} \label{limitation:new_family}
\updated{For any malware detection tool, it is no doubt that detecting new malware families in practice is a very important task. However, neither dynamic nor static methods can fully guarantee the validity of protection against the new malware samples. For example, due to the limited training dataset, MobiTive would have a similar limitation as other static analysis based malware detection systems, which is different from the dynamic analysis approaches. Specifically, considering a new malware family, the situation may be that the malicious features are totally different from existing data. Consequently, as a result of lack of knowledge, the trained classifier may not be able to make the right decision, although learning-based approaches sometimes have the ability to detect new malware variants.
Therefore, in the future, we can make some efforts to improve the ability to detect new samples by combining varies techniques. We surely will also try to improve the ability to detect new malware by designing new adaptive methods, which will also benefit the community by discovering the possible techniques on solving this open question.}

\noindent \textbf{Against adversarial attack.}\label{limitation:adv}
\updated{Indeed, deep learning based systems (e.g., voice/image recognition) will suffer from adversarial attacks~\cite{papernot2016distillation, papernot2016limitations,chen2019can,chen2019real}, so that maintaining the robustness of deep learning based system becomes a challenging topic. However, there are several differences in the deep learning based systems between malware detection task and voice/image recognition. (1) {First} and most important, unlike voice/image recognition, the adversarial attacks in malware detection cannot break the entire functionality in the applications easily in practice, so that the existing adversarial attacks against malware detection are always generated by manipulating the target malware application with un-triggered code snippets (e.g., dead code) instead of changing real functionalities~\cite{chen2019android}. Although it is able to generate adversarial samples to evade the classifier and achieve a high miss-classification rate, it is impractical so far, because such attack can be easily detected by leveraging other techniques such as static data flow analysis to delete such features that are introduced by adding dead code from attackers. Meanwhile, it is also evidenced by lacking real adversarial malware samples in the existing researches. (2) Secondly, different from malware detection approaches on other system (e.g., Windows/Linux), our approach abstracts the entire Android app with a limited feature list instead of embedding the whole package program, so that the attackers have to manipulate their malware applications with our defined features to bypass MobiTive. In practice, attackers can not obtain the accurate feature list easily. Meanwhile, considering that most of our selected features (i.e., manifest properties and API calls) are defined by the official/trustworthy third-party developers, it is almost impossible to bypass MobiTive as easily as the deep learning based voice/image recognition systems under the restriction of maintaining the functionalities in the malware applications. All in all, adversarial attacks on deep learning based malware have domain-specific challenges compared with image/voice classification, which is also belonging to a new research direction as an open question.}

\noindent \textbf{Dynamic behavior analysis.} \label{limitation:adv}
\updated{According to our knowledge and an in-depth literature review, static analysis acts an important in past and current cyber security research, and the number of research publications on Android malware detection is also larger than dynamic analysis (static analysis~\cite{chin2011analyzing, li2018characterising, deshotels2014droidlegacy, zhang2014semantics, zhongyang2013droidalarm, wu2012droidmat, narayanan2016adaptive, meng2016semantic, arp2014drebin, yuan2016droiddetector, yu2014towards, mclaughlin2017deep, kim2018multimodal, xu2018deeprefiner} vs. dynamic analysis~\cite{yan2012droidscope, enck2014taintdroid, performance_counter, shabtai2012andromaly, schmidt2009monitoring, grace2012riskranker}).
Indeed, on a specific given detection task, dynamic behavior analysis may achieve a more accurate result (e.g., lower false positive) than static analysis, however, there are several limitations, which undertake its applicability on specific scenarios, need to be discussed. (1) First and most important, the available scenarios of dynamic behavior analysis based malware detection systems are more limited, because the high cost on computational resources makes dynamic behavior analysis based systems unable to satisfy users' requirements on performance and energy. For example, using performance counter~\cite{performance_counter} while doing program analysis in malware/bug detection task is widely used. However, unlike traditional windows/linux programs, Android application have a more complicated HCI mechanism. In other word, generating good quality test benchmarks with a good coverage to the corner cases is much more difficult than programs on windows/linux. Assuming we have the ability to obtain the benchmarks, the time cost in generating and executing them will also bring a conflict to the target, which is satisfying user's demand on efficiency. (2) Second, the detection efficiency is highly depend on the coverage of the predefined behaviors. Namely, once the malicious behavior in the target malware is not specifically defined by the detection system, the security of system will be no longer promised. (3) Third, different from MobiTive, dynamic behavior analysis based system may suffer from its working mechanism (i.e., before installation vs. run-time). For example, a social engineering based spyware can easily store the privacy information on the device and trick the user to upload them, as a result of that most users are not as professional as security researchers. In the end, according to the diverse usage scenarios and targets, we think Android malware detection approaches based on dynamic behavior and static analysis have their own advantages and weaknesses respectively, which both call for research on them.}

\section{Related Work} \label{related}

\sen{Some techniques are proposed based on analyzing the XML files from the APK file. C.-Y. Huang et al.~\cite{huang2013performance} classified the benign data and malware data using the permission information in manifest and files structure as features. 
Similarly, Z. Aung et al.~\cite{aung2013permission} also considered the permission. 
Differently, they concentrate on the permission requests in the source code, not only the static information. 
E. Chin et al.~\cite{chin2011analyzing} proposed ComDroid, which detects malware by analyzing the manifest file.
There are also techniques which are based on the API \cite{li2018characterising}.
L. Deshotels et al.~\cite{deshotels2014droidlegacy} classified malware based on the API call frequency. 
M. Zhang et al.~\cite{zhang2014semantics} developed DroidSIFT based on the API dependency graphs.
Y. Zhongyang et al.~\cite{zhongyang2013droidalarm} introduced DroidAlarm, which analyzes the inter-procedural call graphs constructed by the relationship between permissions and the interface to identify attacks. 
L. K. Yan et al.~\cite{yan2012droidscope} proposed DroidScope, which generates semantic information from API call and Dalvik opcode traces. 
D.-J. Wu et al.~\cite{wu2012droidmat} proposed the DroidMat to detect malware with API traces, intent, communication and some other life-cycle information.}

\sen{Another line of malware research is conducted based on the program analysis (e.g., control flow graph), which is more expensive than the XML-based and API-based approach. However, the result tends to be more precise. 
Narayanan et al.~\cite{narayanan2016adaptive} presented an online SVM classifier, which uses the control flow graph generated from the source code as input. 
W. Enck et al.~\cite{enck2014taintdroid} proposed TaintDroid, which is a taint analysis tool for Android apps. 
It detects the leakages with the data flow analysis on target sensitive data. 
G. Z. Meng et al.~\cite{meng2016semantic} proposed a deterministic symbolic automaton (DSA) based detection system, in which DSA contains the corresponding components of the target app. 
Furthermore, they developed a system, DroidEcho, which detects attacks with the inter-component communication graphs (ICCG). ICCG provides both the call graphs and sensitive data flow in apps.}
 
\sen{Machine learning has achieved great success in malware detection, there exist also a lot of learning-based approaches.
D. Arp et al.~\cite{arp2014drebin} proposed Drebin, which is a classifier using features from both of XML files and API calls. 
Z. Yuan\cite{yuan2016droiddetector} et al. provided Droid-detector, which performs on a deep belief network. 
W. Yu\cite{yu2014towards} et al. presented a malware detection system, which uses permission and API call traces as input. 
N. McLaughlin\cite{mclaughlin2017deep} et al. used the convolution neural network in detection. The raw opcode sequences of target apps are used as the input feature. Kim\cite{kim2018multimodal} et al. presented a malware detection framework based on multiple neural networks. 
Every network has a single feature input and output score. The final detection result is a combination of all the models. 
K. Xu\cite{xu2018deeprefiner} et al. proposed DeepRefiner, which is an efficient two layer malware detection system. 
They involved XML features as the first layer to perform a fast detection first. 
At the end of the first layer, if it cannot promise the result with a high rate, it will use some more complicated features, like bytecode information, etc., in the second layer to determine whether the target is a malware.} 

\sen{In addition, there are still some other techniques. 
A. Demontis et al.~\cite{demontis2017yes} proposed an algorithm to mitigates attacks like malware data manipulation. 
T. Blsing et al.~\cite{blasing2010android} introduced AASandbox, which performs detection with combination information of both static and dynamic analysis. A. Shabtai et al.~\cite{shabtai2012andromaly} and A.-D. Schmidt et al.~\cite{schmidt2009monitoring} provided the abnormalities identification systems, which use run-time device information, such as CPU usage etc. 
J. Sun et al.~\cite{sun2017malware} trained a machine learning based classifier, which use the distance of keywords to detect the malware. 
L. Lu et al.~\cite{lu2012chex}, P. P. F. Chan et al.~\cite{chan2012droidchecker}, K. Lu et al.~\cite{lu2015checking} and F. Wei et al.~\cite{wei2014amAndroid} focused on detecting vulnerable components, which may hijack the apps. W. Zhou et al.~\cite{hao2014puma} provided a malware detection system, DroidModss, which uses hash comparison to detect repacked Apks. 
M. Grace et al.~\cite{grace2012riskranker} proposed RiskRanker, which performs detection via analyzing specific app behaviors.}

\sen{Existing techniques mainly focused on detecting malware with the information from APK or the source code on server. However, with the rapid development of AI chips on devices, the research about malware detection on mobile side is still rare and on demand. Different from the existing techniques, MobiTive concentrates on using deep learning algorithms on malware detection according to various performance-based experiments on the Android mobile devices.}

\section{Conclusion}\label{conclusion}
\sen{This paper presents MobiTive, a performance-sensitive Android malware detection system on mobile devices as a pre-installed solution.
According to the effectiveness of selected features and the efficiency of feature extraction, MobiTive can provide a reliable detection accuracy and fast responsive (i.e., less than 3 seconds on average) detection service on mobile devices directly. 
To validate the efficiency and reliability, we evaluate MobiTive on six real mobile devices.
To provide more insights of this work, we also make an in-depth analysis of the performance trend on over one hundred mobile phones.}

\section*{Acknowledgments}
This work was supported by Singapore Ministry of Education Academic Research Fund Tier 1 (Award No. 2018-T1-002-069), the National Research Foundation, Prime Ministers Office, Singapore under its National Cybersecurity R\&D Program (Award No. NRF2018 NCR-NCR005-0001), the Singapore National Research Foundation under NCR Award Number NSOE003-0001, NRF Investigatorship NRFI06-2020-0022, the National Research Foundation, Prime Ministers Office, Singapore under NCR Award Number NRF2018NCR-NSOE004-0001, the National Natural Science Foundation of China (No. 61902395). We gratefully acknowledge the support of NVIDIA AI Tech Center (NVAITC).

\ifCLASSOPTIONcaptionsoff
\fi

\bibliographystyle{IEEEtran}
\bibliography{ref}

\begin{IEEEbiography}[{\includegraphics[width=1in,height=1.25in,clip,keepaspectratio]{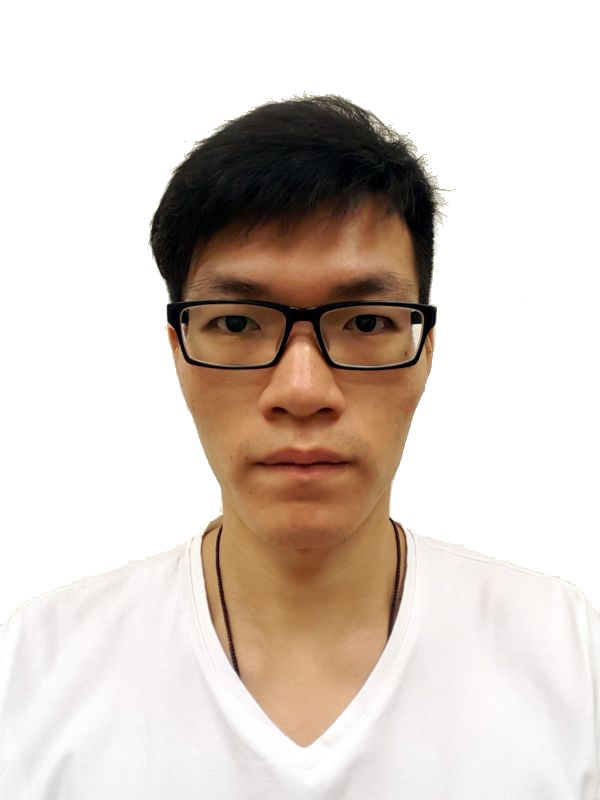}}]
    {Ruitao Feng} received the bachelor’s degree in conputer science and technology from Tianjin University, Tianjin, China, in 2014. He is currently pursuing the Ph.D. degree with the School of Computer Science and Engineering, Nanyang Technological University, Singapore.
    He also works as research assistant in Temasek lab, Nanyang Technological University, Singapore, since 2014. His research interests include solving security and performance problems on mobile platform with software engineering and machine(deep) learning methods.
\end{IEEEbiography}

\vspace{-1cm}

\begin{IEEEbiography}[{\includegraphics[width=1in,height=1.25in,clip,keepaspectratio]{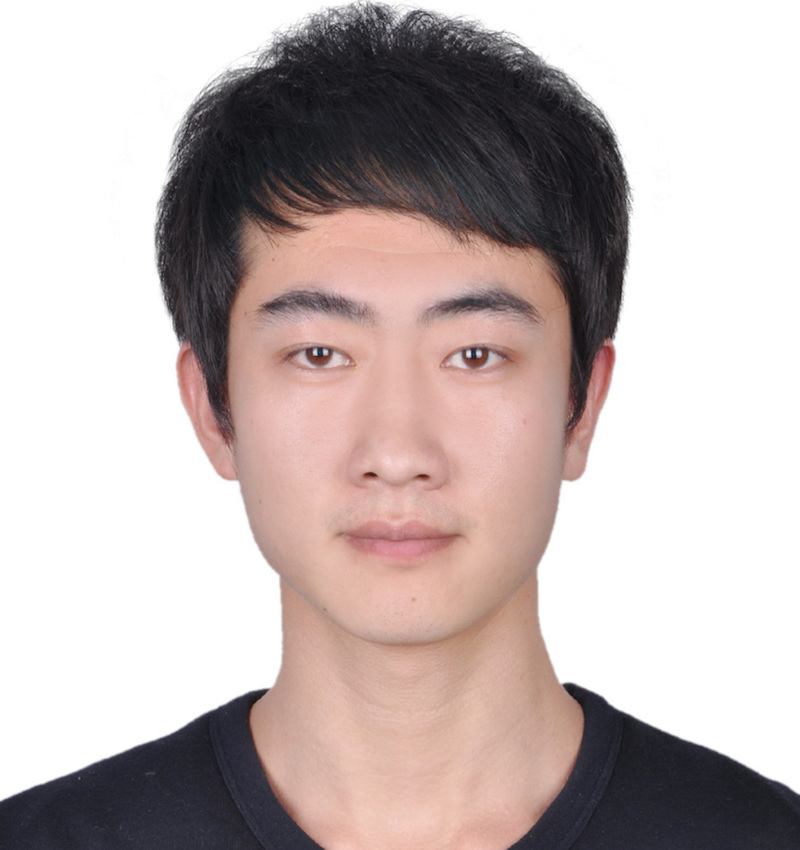}}]
{Sen Chen} received his Ph.D. degree in Computer Science from School of Computer Science and Software Engineering, East China Normal University, China, in June 2019.
Currently, he is a Research Assistant Professor in School of Computer Science and Engineering, Nanyang Technological University, Singapore.
Previously, he was a Research Assistant of NTU from 2016 to 2019 and a Research Fellow from 2019-2020. 
His research focuses on Security and Software Engineering.

\end{IEEEbiography}

\vspace{-1cm}

\begin{IEEEbiography}[{\includegraphics[width=1in,height=1.25in,clip,keepaspectratio]{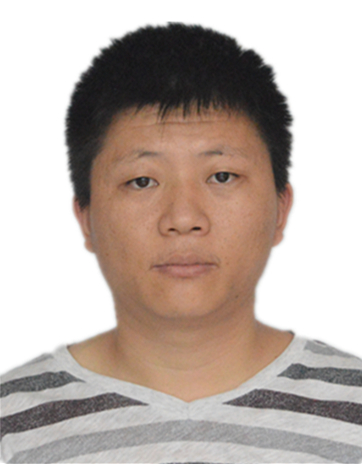}}]
    {Xiaofei Xie} is a presidential postdoctorial fellow in Nanyang Technological University, Singapore. He received Ph.D, M.E. and B.E. from Tianjin University.
    His research mainly focus on program analysis, loop analysis, traditional software testing and security analysis of artificial intelligence.
    He has published some top tier conference/journal papers relevant to software analysis in ISSTA, FSE, TSE, IJCAI and CCS. In particular, he won two ACM SIGSOFT Distinguished Paper Awards.
\end{IEEEbiography}

\vspace{-1cm}

\begin{IEEEbiography}[{\includegraphics[width=1in,height=1.25in,clip,keepaspectratio]{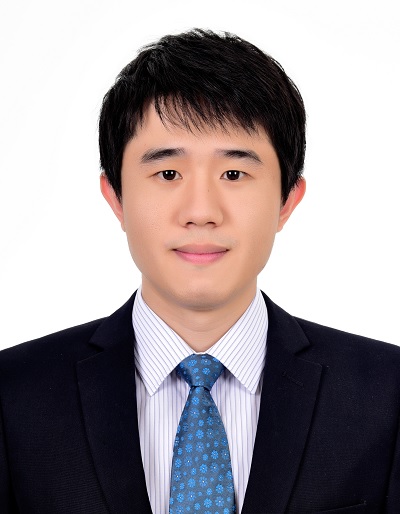}}]
    {Guozhu Meng} received the B.E. and M.E. degrees from Tianjin University, China, in 2009 and 2012, respectively, and the Ph.D. degree from Nanyang Technological University, Singapore, in 2017.
    He is currently an Associate Professor with the Institute of Information Engineering, Chinese Academy of Sciences. Before that, he was a Research Fellow with Nanyang Technological University and a Visiting Research Fellow at the University of Luxembourg. His research interests include mobile security, vulnerability detection, big data analysis.
\end{IEEEbiography}

\vspace{-1cm}

\begin{IEEEbiography}[{\includegraphics[width=1in,height=1.25in,clip,keepaspectratio]{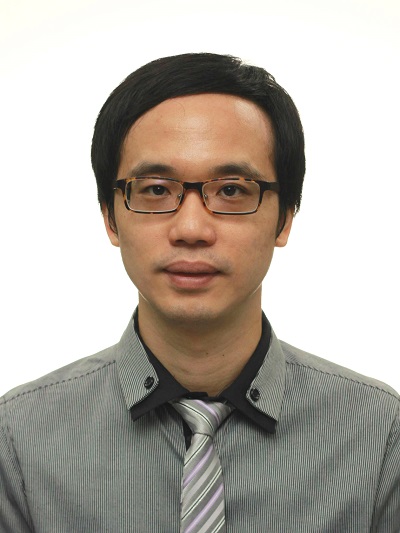}}]
    {Shang-Wei Lin} received his B.S. degree from the National Chung Cheng University in 2003 and his Ph.D. degree in 2010.
    He started his post-doctoral work at NUS and SUTD from 2011 to 2015.
    In May 2015, he joined Nanyang Technological University (NTU) as Assistant Professor. 
    His research interests include formal verification, formal synthesis, embedded system design, cyberphysical systems, security systems, multi-core programming, and component-based object-oriented app frameworks for real-time embedded systems. 
\end{IEEEbiography}

\vspace{-1cm}

\begin{IEEEbiography}[{\includegraphics[width=1in,height=1.25in,clip,keepaspectratio]{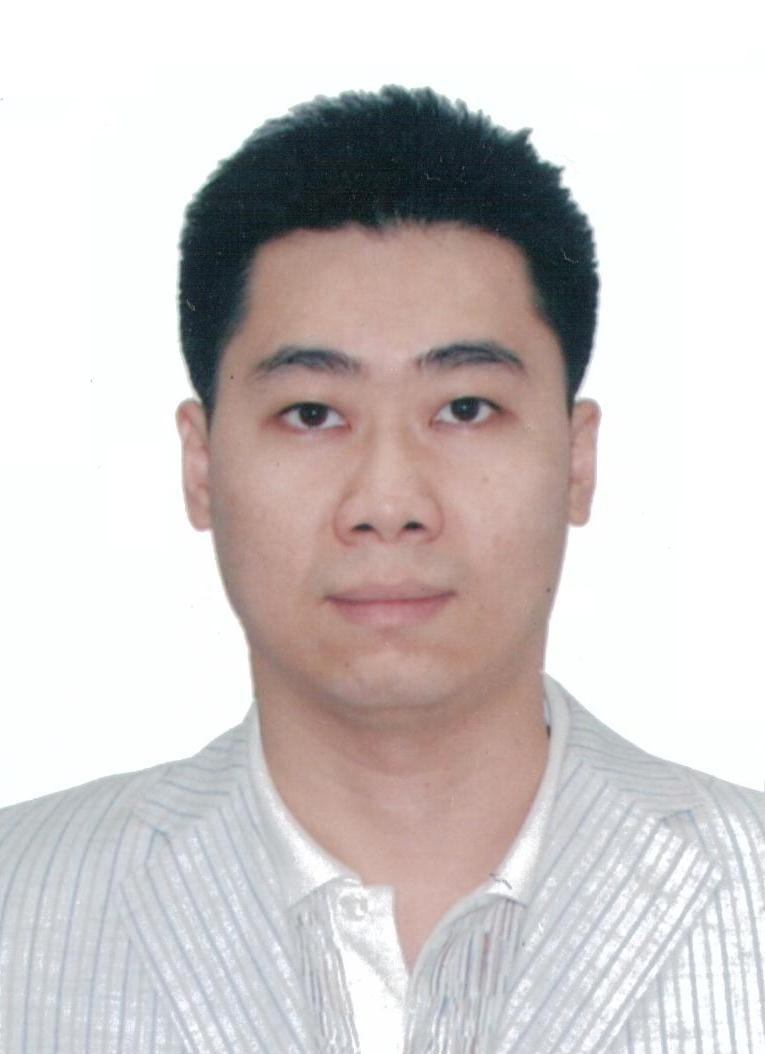}}]
    {Yang Liu} received the Bachelor of Computing degree (Hons.) from the National University of Singapore (NUS) in 2005 and the Ph.D. degree in 2010.
    He started his post-doctoral work at NUS, MIT, and SUTD. In Fall 2012, he joined Nanyang Technological University (NTU) as a Nanyang Assistant Professor. He is currently a Professor and the Director of the Cybersecurity Lab, NTU. He specializes in software verification, security, and software engineering.
\end{IEEEbiography}







\end{document}